\documentclass[final,5p,times,twocolumn]{elsarticle}
\usepackage[utf8]{inputenc}
\usepackage[T1]{fontenc}
\usepackage{url}
\usepackage{booktabs}
\usepackage{colortbl}
\usepackage{subfig} 

\newcommand\ROBAST{\texttt{ROBAST}}
\newcommand\Python{\texttt{Python}}
\newcommand\OpenGL{\texttt{OpenGL}}
\newcommand\matplotlib{\texttt{matplotlib}}
\newcommand\Zemax{\texttt{Zemax OpticStudio}}
\newcommand\simtelarray{\texttt{sim\_telarray}}

\newcommand\CORSIKA{\texttt{CORSIKA}}
\newcommand\Geant{\texttt{Geant4}}
\newcommand\ROOT{\texttt{ROOT}}
\newcommand\PyROOT{\texttt{PyROOT}}
\newcommand\libGeom{\texttt{libGeom}}
\newcommand\cm{\checkmark}
\newcommand\TGraph{\texttt{TGraph}}
\newcommand\TGeoSphere{\texttt{TGeoSphere}}
\newcommand\TGeoBBox{\texttt{TGeoBBox}}
\newcommand\TGeoPgon{\texttt{TGeoPgon}}
\newcommand\TGeoManager{\texttt{TGeoManager}}
\newcommand\AOpticsManager{\texttt{AOpticsManager}}
\newcommand\AMirror{\texttt{AMirror}}
\newcommand\ALens{\texttt{ALens}}
\newcommand\AGeoAsphericDisk{\texttt{AGeoAsphericDisk}}
\newcommand\AGeoWinstonConePoly{\texttt{AGeoWinstonConePoly}}
\newcommand\AGeoBezierPgon{\texttt{AGeoBezierPgon}}
\newcommand\ARay{\texttt{ARay}}
\newcommand\AObscuration{\texttt{AObscuration}}
\newcommand\AFocalSurface{\texttt{AFocalSurface}}
\newcommand\AOpticalComponent{\texttt{AOpticalComponent}}
\newcommand\ACorsikaIACTFile{\texttt{ACorsikaIACTFile}}
\newcommand\ARefractiveIndex{\texttt{ARefractiveIndex}}
\newcommand\CPP{\texttt{C++}}
\newcommand\CARE{\texttt{CARE}}
\newcommand\GrOptics{\texttt{GrOptics}}
\newcommand\git{\texttt{git}}
\newcommand\Bezier{B\'{e}zier}

\definecolor{Gray}{gray}{0.95}

\usepackage{graphicx}

\usepackage{amssymb}

\usepackage{multirow}

\biboptions{}

\journal{Astroparticle Physics}

\begin{document}

\begin{frontmatter}



\title{ROBAST: Development of a ROOT-Based Ray-Tracing Library for Cosmic-Ray Telescopes and its Applications in the Cherenkov Telescope Array}


\author[Nagoya,MPIK,Leicester]{Akira Okumura\corref{cor}}
\ead{oxon@mac.com}
\author[MPI]{Koji Noda}
\author[Minnesota,Paris]{Cameron Rulten}

\cortext[cor]{Correspondence and software support}

\address[Nagoya]{Institute for Space-Earth Environmental Research, Nagoya University, Furo-cho, Chikusa-ku, Nagoya, Aichi 464-8601, Japan}
\address[MPIK]{Max-Planck-Institut f\"{u}r Kernphysik, P.O. Box 103980, D 69029 Heidelberg, Germany}
\address[Leicester]{Formerly at Department of Physics and Astronomy, University of Leicester, University Road, Leicester, LE1 7RH, UK}
\address[MPI]{Max-Planck-Institut f\"{u}r Physik, F\"{ö}hringer Ring 6, D 80805 M\"{u}nchen, Germany}
\address[Minnesota]{Department of Physics and Astronomy, University of Minnesota, 116 Church Street, Minneapolis, MN, 55455, U.S.A.}
\address[Paris]{Formerly at LUTH, Observatoire de Paris, CNRS, Universite Paris Diderot, 5 Place Jules Janssen, 92190 Meudon, France}

\begin{abstract}
We have developed a non-sequential ray-tracing simulation library, \texttt{ROOT-based} \texttt{simulator} \texttt{for} \texttt{ray} \texttt{tracing} (\ROBAST), which is aimed to be widely used in optical simulations of cosmic-ray (CR) and gamma-ray telescopes. The library is written in \CPP, and fully utilizes the geometry library of the \ROOT\ framework. Despite the importance of optics simulations in CR experiments, no open-source software for ray-tracing simulations that can be widely used in the community has existed. To reduce the dispensable effort needed to develop multiple ray-tracing simulators by different research groups, we have successfully used \ROBAST\ for many years to perform optics simulations for the Cherenkov Telescope Array (CTA). Among the six proposed telescope designs for CTA, \ROBAST\ is currently used for three telescopes: a Schwarzschild--Couder (SC) medium-sized telescope, one of SC small-sized telescopes, and a large-sized telescope (LST). \ROBAST\ is also used for the simulation and development of hexagonal light concentrators proposed for the LST focal plane. Making full use of the \ROOT\ geometry library with additional \ROBAST\ classes, we are able to build the complex optics geometries typically used in CR experiments and ground-based gamma-ray telescopes. We introduce \ROBAST\ and its features developed for CR experiments, and show several successful applications for CTA.
\end{abstract}


\end{frontmatter}


\section{Introduction}
\label{sec:intro}

The detection sensitivity in observations of very-high-energy (VHE) gamma rays and ultra-high energy (UHE) cosmic rays (CRs) crucially depends on the optical system. In UHE CR telescopes such as the Telescope Array and the Pierre Auger Observatory, the air fluorescence emission induced by extensive air showers is observed by reflective telescopes with wide field-of-views (FOVs) ($15^\circ$--$30^\circ$) and large diameters (${\sim}3$~m) \cite{Tokuno:2009:On-site-calibration-for-new-fluorescence-detectors,Abraham:2010:The-fluorescence-detector-of-the-Pierre-Auger-Obse} to image long air-shower trajectories across the sky. In contrast, VHE gamma-ray telescopes, which are usually designed with narrower FOVs ($3^\circ$--$5^\circ$) and larger-diameter mirrors (${\sim}10$--$30$~m)  \cite{Bernlohr:2003:The-optical-system-of-the-H.E.S.S.-imagi,Holder:2006:The-first-VERITAS-telescope}, observe the atmospheric Cherenkov radiation yielded by electromagnetic air showers to sensitively detect gamma rays at a threshold of as low as several tens of GeV.

The design, performance evaluation, tolerance analysis, and Monte Carlo event simulations of these optical systems are frequently based on the ray-tracing technique, which calculates individual photon tracks using dedicated software. For instance, the fluorescence detectors of the Pierre Auger Observatory \cite{Abraham:2010:The-fluorescence-detector-of-the-Pierre-Auger-Obse} are simulated with the optical photon processes in \Geant\ \cite{Agostinelli:2003:Geant4--a-simulation-toolkit}, and \simtelarray\ \cite{Bernlohr:2008:Simulation-of-imaging-atmospheric-Cherenkov-telesc} has been developed for the HEGRA IACT system \cite{Daum:1997:First-results-on-the-performance-of-the-HEGRA-IACT}, the H.E.S.S. telescopes \cite{Bernlohr:2003:The-optical-system-of-the-H.E.S.S.-imagi}, and the Cherenkov Telescope Array (CTA) \cite{Bernlohr:2013:Monte-Carlo-design-studies-for-the-Cherenkov-Teles}. In addition, various groups and institutions have developed their own ray-tracing programs (e.g., \cite{Sasaki:2002:Design-of-UHECR-telescope-with,Lopez:2013:Simulations-of-the-MAGIC-telescopes-with-matelsim}), whereas others have used commercial software such as \Zemax\footnote{\url{http://www.zemax.com/}} (e.g., \cite{Zech:2013:SST-GATE:-A-dual-mirror-telescope-for-the-Cherenko,Aguilar:2015:Design-optimization-and-characterization-of-the-li}).

Despite the wide variety of ray-tracing programs used in gamma-ray and CR research, a standard open-source program that can be widely disseminated among different research groups is currently lacking. On the other hand, many optical simulations of Cherenkov or fluorescence telescopes (hereafter collectively CR telescopes) require the same or similar software, even in different projects. Thus, developing a useful and accurate ray-tracing program that is freely available in the community should not only improve the flexibility of simulations but also reduce the effort and time expended in parallel software development, debugging, and verification processes.

The present paper introduces an open-source ray-tracing library compatible with diverse optical CR telescope studies. The functionality of our software is demonstrated in practical applications to CTA.

\begin{table*}
\centering
\caption{A comparison of ray-tracing programs. Only functionality that is important for CR telescopes is listed.}
\label{tab:comparison}
\begin{tabular}{lccccc}
\toprule
& \ROBAST & \multicolumn{2}{c}{\Zemax} & \simtelarray & \Geant \\
& & {\small Professional} & {\small Standard} & & \\
\rowcolor{Gray}Non-sequential ray tracing & \cm & \cm & & & \cm \\
Refraction & \cm & \cm & \cm & & \cm \\
\rowcolor{Gray}Diffraction & & \cm & \cm & & \\
Polarization & $^\star$ & \cm & \cm & & \cm \\
\rowcolor{Gray}Composite 3D Objects & \cm & \cm & & & \cm \\
Aspherical mirrors/lenses & \cm & \cm & \cm & \cm$^\ast$ & \\
\rowcolor{Gray}Winston cones & \cm & \cm & & & \\
CAD import & & \cm & & & $^\mathrm{\,\,\,}$\cm$^\dagger$ \\
\rowcolor{Gray}\CPP & \cm & $^\mathrm{\,\,\,}$\cm$^\ddagger$ & & \cm & \cm \\
\Python\ & \cm & $^\mathrm{\,\,\,}$\cm$^\ddagger$ & & & \cm \\
\rowcolor{Gray}\texttt{CORSIKA IACT} interface & \cm & & & \cm & \\
OS X & \cm & & & & \cm \\
\rowcolor{Gray}Linux & \cm & & & \cm & \cm \\
3D visualization & \cm & \cm & \cm & & \cm \\
\rowcolor{Gray}Optimization engine & $^\mathrm{\,\,\,}$\cm$^\P$ & \cm & \cm & & \\
& Open Source & & & Open Source & Open Source\\
\multirow{-2}{*}{License} & (LGPL$^\S$) & \multirow{-2}{*}{Commercial} & \multirow{-2}{*}{Commercial} & (GPL$^\S$) & (Geant4 Software License)\\
\bottomrule
\multicolumn{6}{l}{$^\star$ {\small Polarization is not currently supported, but we have a plan to support it in a future \ROBAST\ release.}} \\
\multicolumn{6}{l}{$^\ast$ {\small Lens simulation is not supported.}} \\
\multicolumn{6}{l}{$^\dagger$ {\small Requires an indirect conversion from a CAD STEP file to a GDML file using external software.}} \\
\multicolumn{6}{l}{$^\ddagger$ {\small Available as external \Zemax\ extensions or through the Dynamic Data Exchange protocol on Windows.}} \\
\multicolumn{6}{p{40em}}{$^\P$ {\small Various optimization engines provided with \ROOT\ can be used to optimize optical system parameters, however the user needs to write some \CPP\ code even for a very simple optical system.}} \\
\multicolumn{6}{l}{$^\S$ {\small GNU Lesser General Public License (LGPL) and General Public License (GPL)}}
\end{tabular}
\end{table*}

\section{ROOT-based Simulator for Ray Tracing}
\label{sec:ROBAST}

We have developed a ray-tracing simulation library, \texttt{ROOT-based} \texttt{simulator} \texttt{for} \texttt{ray} \texttt{tracing} (\ROBAST), which utilizes the geometry library (\libGeom) of \ROOT\footnote{\url{http://root.cern.ch/}}. \ROBAST\ has been developed as an open-source library, and the source code and online documentation are publicly available from the \ROBAST\ \git\ repository\footnote{\url{https://github.com/ROBAST/ROBAST/}} and the \ROBAST\ website\footnote{\url{https://robast.github.io/}}, respectively. Many tutorial programs are also provided with the library\footnote{ROBAST example programs used for Figures~\ref{fig:Ashra}, \ref{fig:HESS}, \ref{fig:spider}, and \ref{fig:Okumura} are also provided as tutorials.}.

\subsection{\ROOT\ and \libGeom}

The data analysis framework \ROOT\ is extensively employed in high-energy particle physics and astroparticle physics \cite{Brun:1997:ROOT----An-object-oriented-data-analysis}. In addition to the analysis and mathematical libraries (written in \CPP), \ROOT\ contains a geometry library (\libGeom) that provides a range of functionalities for building, browsing, tracking, and visualizing detector geometries in high-energy particle experiments \cite{Brun:2003:The-ROOT-geometry-package}. \libGeom\ can track particles with arbitrary position and momentum vectors, moving through detector geometries. When a moving particle crosses the boundary surface of a detector geometry, \libGeom\ calculates the coordinates of the intersection and the vector normal to the boundary. Hence, optical reflections and refractions on media boundaries are easily simulated by adding dedicated classes to \libGeom.

The imaging systems in CR experiments must be resolved to the size of intrinsic air-showers (typically of order $0.1^\circ$ to $1^\circ$). Therefore, a geometrical optics approach is adequate for simulating most CR telescopes, and the particle tracking method in \libGeom\ is applicable to most cases.

\subsection{\ROBAST\ Features and Software Requirements in Ray-Tracing Simulations of CR Telescopes}
\label{subsec:requirements}

The development of \ROBAST\ was motivated by identifying several common software requirements in ray-tracing simulations of CR telescopes. All these requirements cannot be covered by a single existing method such as \Zemax, \simtelarray, or \Geant\ (Section \ref{subsec:comparison}). However, utilizing the \ROOT\ \libGeom, these requirements are addressed by the \ROBAST\ features discussed in this section. Table~\ref{tab:comparison} compares the \ROBAST\ functionalities with those of other programs.

\subsubsection{Non-Sequential Ray Tracing}

A simple optical system, e.g., an optical system comprising a single parabolic mirror and a focal plane, is commonly simulated by the sequential ray-tracing method, wherein the user specifies the order of all optical component surfaces that reflect, refract, or absorb photons. The simulated photons sequentially reach the media boundaries in the given order.

In contrast, more complex optical simulations involve Fresnel reflections, scattering, multiple reflections on segmented mirror facets, and obscuration by mechanical structures. Such scenarios, wherein the order of the photon-visited surfaces cannot usually be known in advance, are instead simulated by the non-sequential ray-tracing method. As its name indicates, this method calculates the photon tracks connected between several points on the surface boundaries without requiring the surface order in advance.

The reflectors in CR telescopes are mostly constructed from segmented mirror facets to realize large mirror apertures with less expensive technologies. In addition, to simultaneously reduce the dead space and increase the photon collection efficiency, light concentrators are often installed in front of the photodetectors. However, the sequential methods cannot easily simulate multiple reflections inside a light concentrator. Moreover, shadowing by the telescope structure needs to be accurately estimated to reduce the systematic uncertainty in the effective area of the optical system. These situations are adequately handled by the non-sequential method.

By adopting \libGeom\ as the photon tracking engine of \ROBAST, we are able to provide non-sequential ray-tracing functionality, whereas \simtelarray\ and the standard version of \Zemax\ provide only the sequential method as shown in Table~\ref{tab:comparison}\footnote{\simtelarray\ supports a non-sequential mode only in simulations of shadowing by the telescope structure.}. Thus, complex CR telescope geometries cannot be simulated with these programs.

\subsubsection{Simulation of Lenses and Refraction}

Some optical systems in recent CR experiments are installed with corrector lenses that simultaneously widen the FOV and improve the angular resolution. For example, the Ashra experiment uses a modified Baker--Nunn optical system installed with three aspheric-plano lenses made of UV transparent acrylic plates \cite{Aita:2008:Ashra-Mauna-Loa-Observatory-an}. Segmented corrector lenses and Fresnel lenses are used in the fluorescence telescopes of the Pierre Auger Observatory \cite{Abraham:2010:The-fluorescence-detector-of-the-Pierre-Auger-Obse} and the optical system of JEM-EUSO \cite{Marchi:2010:The-JEM-EUSO-Mission-and-Its-Challenging-Optics-Sy}, respectively. The FACT telescope employs light concentrators that utilize total internal reflection \cite{Anderhub:2013:Design-and-operation-of-FACT--the-first-G-APD-Cher}. To simulate these optical systems, the ray-tracing program must account for the refractive effects of lenses as well as the reflective effects of mirrors.

\ROBAST\ calculates the reflection and refraction angles at media boundaries using the normal and momentum vectors of photons computed by \libGeom. These calculations are performed by \AOpticsManager, \AMirror, and \ALens\ classes (see Section~\ref{subsec:classes} for details).

Most ray-tracing simulators support simulation of lenses and refraction. But \simtelarray\ does not have a lens simulation functionality because it is dedicated to ground-based gamma-ray telescopes without lens components.

\subsubsection{Flexibility in Geometry Construction}
\label{subsec:flexibility}

Almost all CR telescopes are installed with large-diameter reflecting systems composed of many segmented mirrors. Ground-based gamma-ray telescopes are equipped with Davies--Cotton optical systems \cite{Davies:1957:Design-of-the-Quartermaster-So} or segmented parabolic systems, whose mirrors comprise circular, hexagonal, or square facets with spherical surfaces. In addition, optical systems with segmented or aspherical lenses have been newly proposed and realized in CR experiments \cite{Abraham:2010:The-fluorescence-detector-of-the-Pierre-Auger-Obse,Aita:2008:Ashra-Mauna-Loa-Observatory-an}. Schwarzschild--Couder (SC) optical systems (which comprise aspherical primary and secondary mirrors) are presently being considered for CTA as they simultaneously improve the angular resolution and widen the FOV \cite{Vassiliev:2007:Wide-field-aplanatic-two-mirro,Actis:2011:Design-concepts-for-the-Cherenkov-Telesc}. Consequently, functionalities that construct complex geometries in optics simulations are increasingly in demand.

\begin{figure}
  \centering
  \includegraphics[width=\columnwidth]{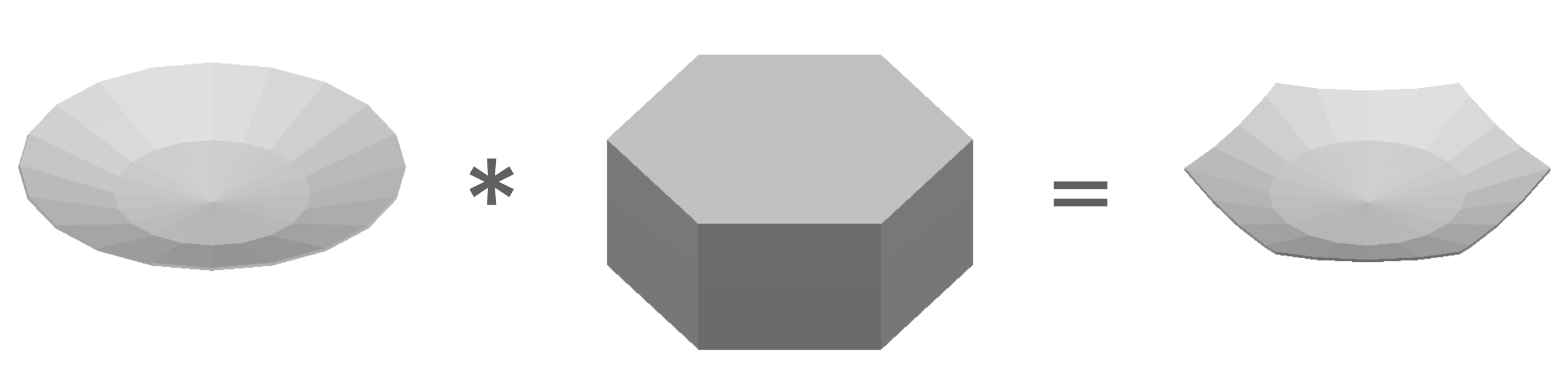}
  \caption{An example of a hexagonal spherical mirror usually installed in CR telescopes. This shape is created from the intersection between a partial sphere created with the \TGeoSphere\ class and a hexagonal prism created with the \TGeoPgon\ class.}
  \label{fig:composite}
\end{figure}

The core engine of \ROBAST, \libGeom, allows the user to build a variety of intricate geometries from more primitive shapes, such as spheres, rectangular boxes, cones, and tubes. Composite shapes can be created from primitive shapes using the Boolean operators union (\texttt{+}), subtraction (\texttt{-}), and intersection (\texttt{*}). A hexagonal spherical mirror created with \ROBAST\ is illustrated in Figure~\ref{fig:composite}.

\begin{figure}
  \centering
  \includegraphics[width=8cm]{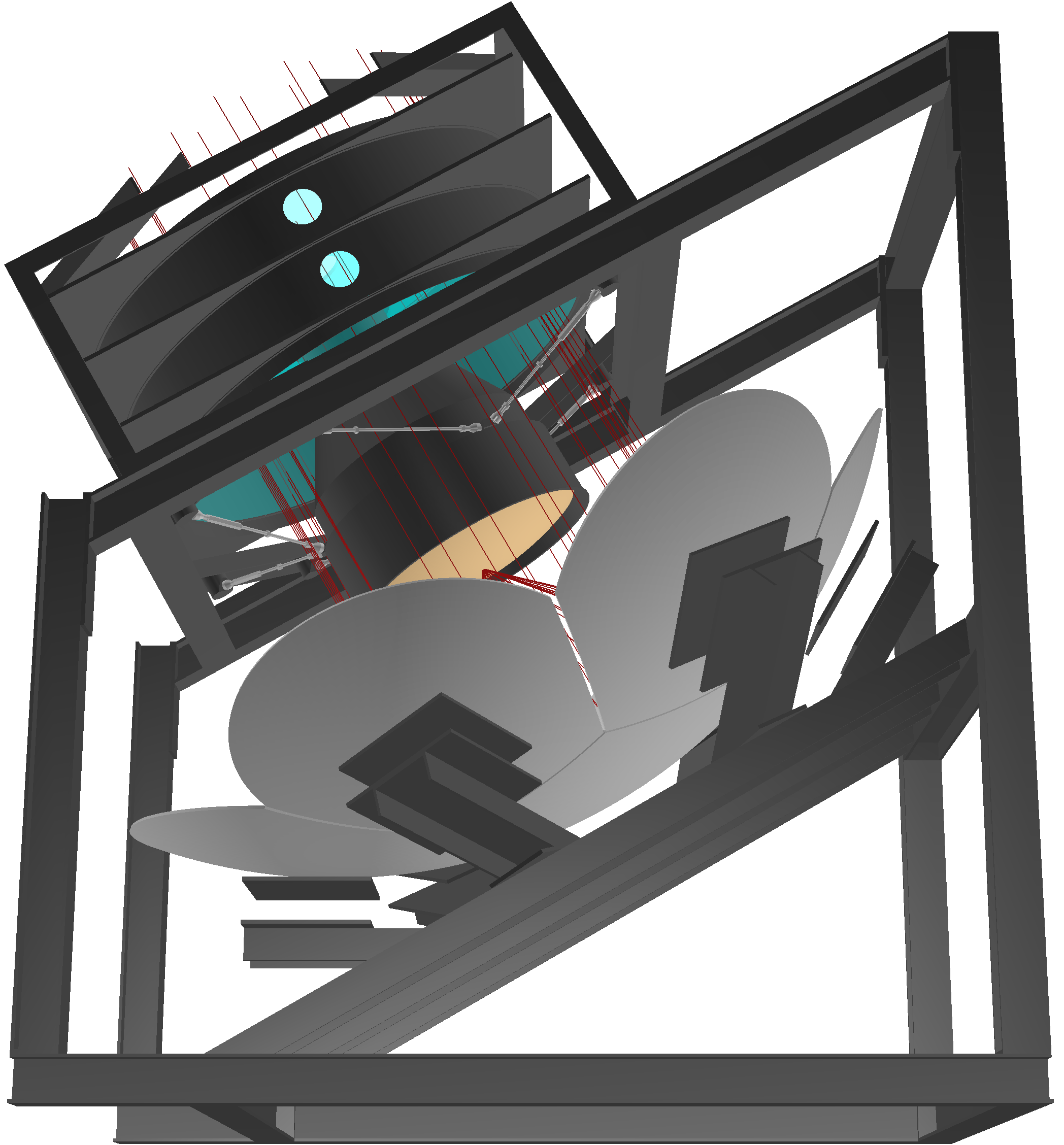}
  \caption{A \ROBAST\ simulation of a modified Baker--Nunn optical system developed for the Ashra experiment \cite{Sasaki:2002:Design-of-UHECR-telescope-with,Aita:2008:Ashra-Mauna-Loa-Observatory-an}. The system comprises three aspherical lenses (light blue in the online version), seven segmented spherical mirrors (light gray), and an image intensifier functioning as a spherical focal surface (dark yellow). The telescope frames (black) are photon absorbers. Red lines show the tracks of simulated photons.}
  \label{fig:Ashra}
\end{figure}

Although many primitive-shape classes exist in \libGeom, these shapes cannot create many important structures of CR telescopes, such as aspherical lenses and mirrors as well as compound parabolic concentrators \cite{Winston:1970:Light-Collection-within-the-Fr,Punch:1994:Winston-Cones-for-Light-Collection-and-Albedo-Prot} (the latter are known as ``Winston cones''). \ROBAST\ can simulate light concentrators and optical systems with aspherical surfaces using additional shape classes (\AGeoWinstonConePoly\ and \AGeoAsphericDisk, respectively). Figure~\ref{fig:Ashra} illustrates a \ROBAST\ simulation of a modified Baker--Nunn optical system comprising three aspherical lenses and segmented spherical mirrors. This complex example exhibits the advantage of \ROBAST\ over other programs (the \Zemax\ standard version, \simtelarray, and \Geant) that cannot simulate optical systems with segmented aspherical shapes or a hexagonal light concentrator in the non-sequential mode. A \ROBAST\ application for a hexagonal light concentrator is given in subsection~\ref{subsec:LG}.

Importing 3D geometries from computer-aided design (CAD) software is sometimes required in simulations of composite telescope structures. While the \Zemax\ professional version and \Geant\ can import 3D geometry files (STL/IGES/STEP/SAT and geometry description markup language (GDML) \cite{Chytracek:2006:Geometry-Description-Markup-Language-for}, respectively), \ROBAST\ is currently unable to import any external files, because \libGeom\ does not fully support the GDML format and thus tessellated objects defined in the GDML format cannot be simulated. If \libGeom\ fully supports GDML in the future, then \ROBAST\ will be able to import 3D CAD geometries using GDML files.

\subsubsection{Connectivity with Other Software}

In simulations of CR telescopes with a ray-tracing program, air-shower simulators, electronics simulators, and data analysis software are often combined. Therefore, we provide \ROBAST\ as a \CPP\ library, enabling users to easily write their own software linked to other programs or libraries. To easily pipe Cherenkov photon data generated by \CORSIKA\ \cite{Heck:1998:CORSIKA:-A-Monte-Carlo-Code-to} to the ray-tracing simulations, \ROBAST\ also provides an interface class (\ACorsikaIACTFile) that reads photon data files within \ROBAST.

Another advantage of \ROBAST\ is seamless connectivity with \ROOT, which is extensively used in Monte Carlo simulations, data acquisition systems, and data analyses in CR experiments. Since \ROBAST\ is written using \ROOT\ libraries, \ROBAST\ simulation results are easily analyzed in the \ROOT\ framework. For example, spot diagrams (also referred to as geometrical point spread functions, PSFs) can be directly plotted using the 2D histogram classes provided in \ROOT. By calling \ROBAST\ classes from \Python\ through \PyROOT, users can combine \ROBAST\ with other \Python\ packages such as \matplotlib.

\subsubsection{Multi-platform Support}
\ROBAST\ is written in standard \texttt{C++03} and requires only the \ROOT\ libraries. It supports the two most used operating systems in the CR community, OS~X and Linux\footnote{\ROOT\ version 5 supports other POSIX and Windows systems. However, due to lack of requests from \ROBAST\ users, \ROBAST\ has not been tested on other platforms.}.

\subsubsection{3D Visualization}

Verifying the optics geometries is important in ray-tracing simulations. If a user's program constructs an incorrect geometry of an optical system, the simulated PSFs will differ from the actual ones, and systematic errors can be generated from the estimated shadowing effect of the telescope masts, camera housing, and secondary mirror.

Without the visualization functionality, verification of ray-tracing simulators becomes more difficult. For instance, to visually check the accuracy of the given mirror geometries and shadowing, the user often draws the mirror shapes and shadows in 3D space by plotting the coordinates of a large number of photons that hit the segmented mirrors. In contrast, using the 3D visualization functionality provided with \libGeom, users can more easily and more directly verify complex optical systems and their geometries visually. As shown in Figure~\ref{fig:Ashra}, all the optical components and simulated rays can be drawn in an \OpenGL\footnote{\url{https://www.opengl.org}} view.

\subsection{Classes}
\label{subsec:classes}

\begin{figure}
  \centering
    \includegraphics[width=0.8\columnwidth]{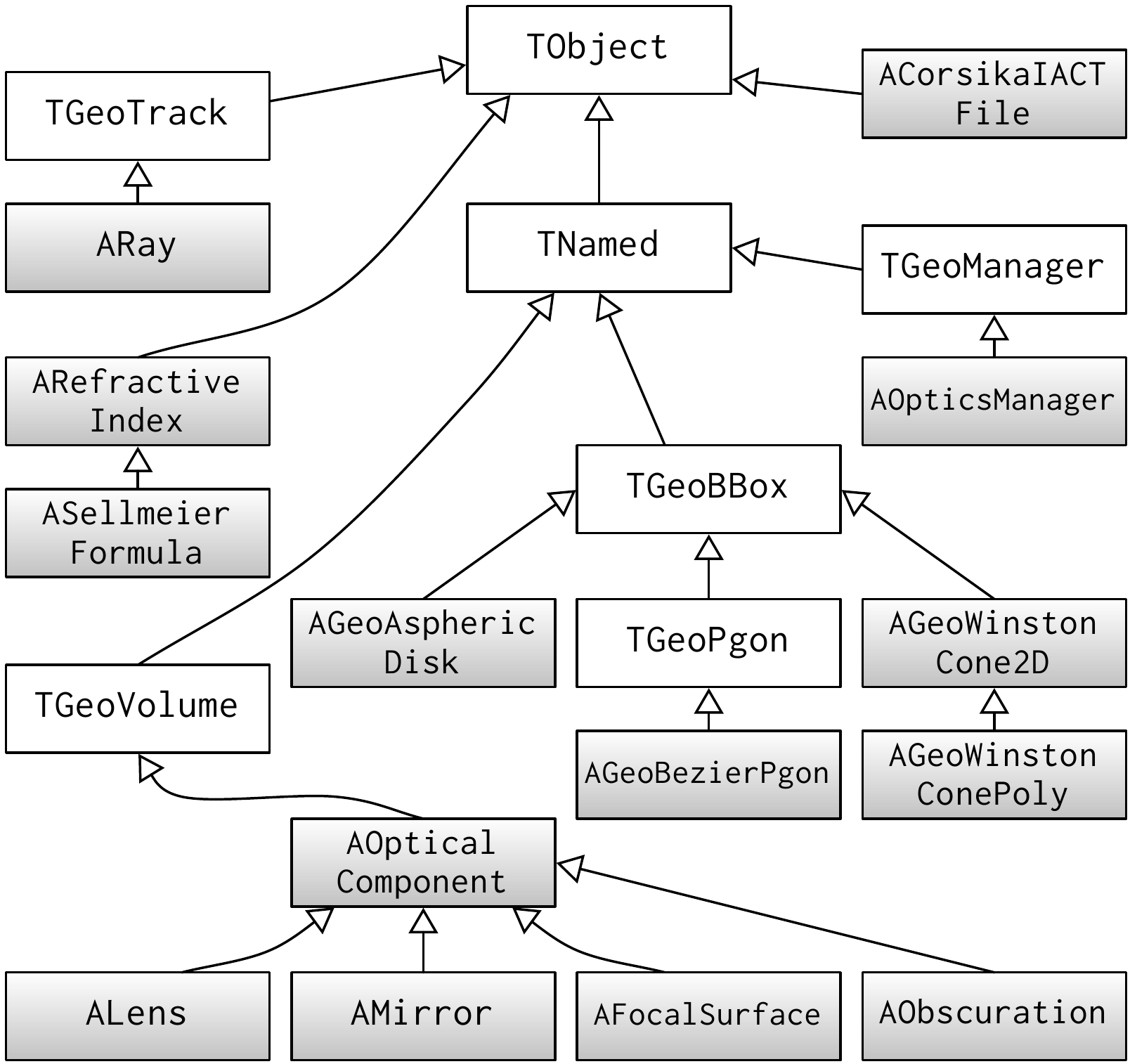}
  \caption{A class diagram of \ROBAST. Class names starting with ``\texttt{T}'' and ``\texttt{A}'' belong to \ROOT\ and \ROBAST, respectively. The latter classes are also shaded. Triangular arrowheads point to the base classes. The inheritance relationships are partially simplified by omitting multiple inheritance and intermediate classes in the relationships. Some of the \ROBAST\ classes are omitted in this diagram.}
  \label{fig:class}
\end{figure}

Although \libGeom\ has excellent particle tracking and geometry construction functionality, it requires additional classes for ray-tracing simulations. \ROBAST\ provides an optics simulation functionality, which is realized by the classes shown in Figure~\ref{fig:class}.

The \TGeoManager\ class (inherited by \AOpticsManager\ class in \ROBAST) manages the particle tracking engine and geometry shapes in \libGeom. \AOpticsManager\ calculates the reflection, refraction, absorption, and photon scattering processes in user-defined optical systems through four distinct classes derived from \AOpticalComponent: \ALens, \AMirror, \AFocalSurface, and \AObscuration.

\begin{itemize}
\item The \ALens\ class\\
Any refractive medium can be modeled with the \ALens\ class. In addition to normal optical lenses, \ALens\ represents optical components with refractive indices or absorption lengths, such as the atmosphere, scintillators, and input windows of the photomultiplier tubes (PMTs). The user specifies the refractive indices using classes derived from \ARefractiveIndex\ and can set the absorption lengths using the \TGraph\ class of \ROOT. Both these parameters are wavelength-dependent.

\item The \AMirror\ class\\
The \AMirror\ class models mirrors. The user specifies the mirror reflectance, which depends on the photon incidence angles and wavelengths. Light concentrators with specular surfaces are also modeled with this class.

\item The \AFocalSurface\ class\\
Photodetectors at the focal planes are modeled with \AFocalSurface. This class has a quantum efficiency (QE) property, which also depends on the photon incidence angles and wavelengths.

\item The \AObscuration\ class\\
A body modeled with the \AObscuration\ class absorbs all incident photons. \AObscuration\ simulates the shadowing caused by telescope frames, masts, and camera housings.
\end{itemize}
The times, coordinates, and momenta of photons are contained in the \ARay\ class. \ARay\ also records all positions where a photon is reflected, refracted, or scattered.

Photon polarization is not yet supported in the current version of \ROBAST\ (version v2.3p1 as of October 2015), and thus the user may need to pay special attention when his/her optical system could propagate photons with large incidence angles at optics boundaries. We have a plan to support photon polarization in a future version.

As mentioned in Section~\ref{subsec:flexibility}, complex geometries of CR telescopes require additional shape classes. Figure~\ref{fig:composite} shows how a simple composite geometry can be constructed from the primitive shapes implemented in \libGeom. However, aspherical lenses, aspherical mirrors, or Winston cones cannot be constructed from such primitive shapes. Therefore, we have implemented \AGeoAsphericDisk\ and \AGeoWinstonConePoly\ classes in order to model these shapes in \ROBAST. We have also implemented the \AGeoBezierPgon\ class that simulates light concentrators composed of \Bezier-curve walls (also referred to as Okumura cones) \cite{Okumura:2012:Optimization-of-the-collection-efficiency-of-a-hex}.

All the classes derived from \AOpticalComponent, along with the shape classes provided by \libGeom\ and \ROBAST, should match the software requirements for the optical systems in most CR telescopes. If the user requires a new shape that is unavailable in \libGeom\ or \ROBAST, they can add new shapes by implementing a new class derived from \TGeoBBox.

The imperfection of optical systems can be also taken into account in \ROBAST\ simulations. For example, profile deviation from ideal mirror surfaces (``form deviation'') and the surface roughness of optics can be simulated (reflection and refraction angles are blurred) by using a boundary roughness property approximated by a 2D Gaussian. Misalignment of individual mirror facets and relocation of the camera position due to gravity can be mathematically described using the translation and rotation matrix classes in \libGeom.

\subsection{Cross-check with Other Programs}

We verified the \ROBAST\ classes and their ray-tracing calculations in comparisons with other independent programs, namely, \simtelarray\ (version 2015-07-21) and \Zemax\ (version 13 release 2). As a cross-check, we here compare the spot diagrams of a Davies--Cotton telescope (H.E.S.S.~I) simulated by \simtelarray\ and \ROBAST, and those of a Schmidt--Cassegrain telescope by \Zemax\ and \ROBAST. Another comparison among these three programs for a CTA telescope is given elsewhere \cite{Armstrong:2015:Monte-Carlo-Studies-of-the-GCT-Telescope-for-the-C,Rulten:2015}.

\begin{figure}
  \centering
  \includegraphics[width=9cm]{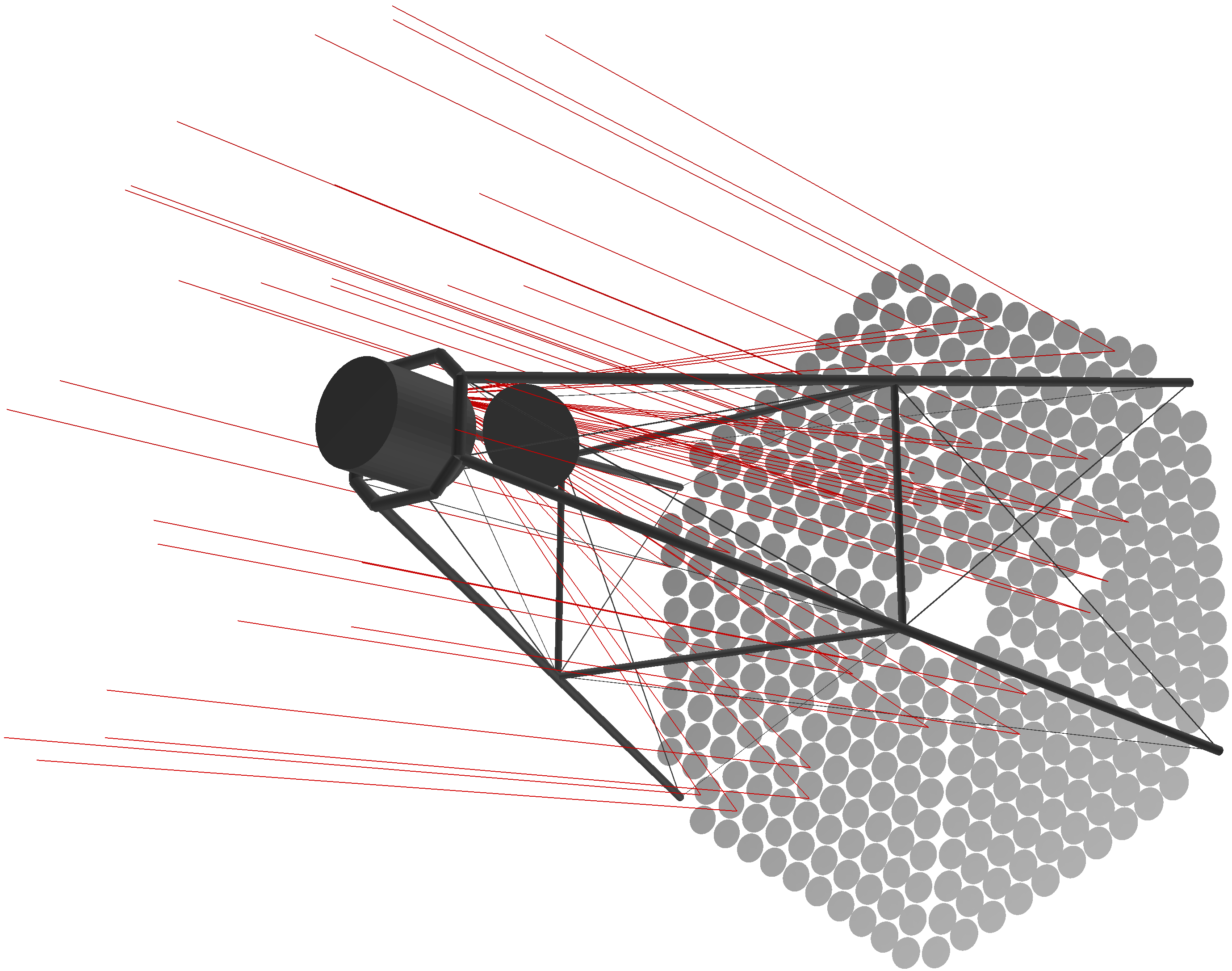}
  \caption{The 3D geometry of an H.E.S.S.~I Davies--Cotton telescope simulated with \ROBAST. The geometry parameters were taken from H.E.S.S. configuration files provided in \simtelarray\ (version 2015-07-21). The system comprises 380 segmented spherical mirror facets, a flat focal plane (not visible in this figure), the camera housing and circular lid, and telescope masts and trusses. Lines show the tracks of the simulated photons. Note that the space of two missing mirrors is used for telescope calibration instrumentation.}
  \label{fig:HESS}
\end{figure}

Figure~\ref{fig:HESS} shows the 3D \ROBAST\ geometry of the Davies--Cotton optical system used in H.E.S.S.~I telescopes. The 380 segmented spherical mirrors and the telescope structures (masts, trusses, camera housing, and camera lid) are modeled with the \AMirror\ and \AObscuration\ classes, respectively.

\begin{figure}
  \centering
  \includegraphics[width=\columnwidth]{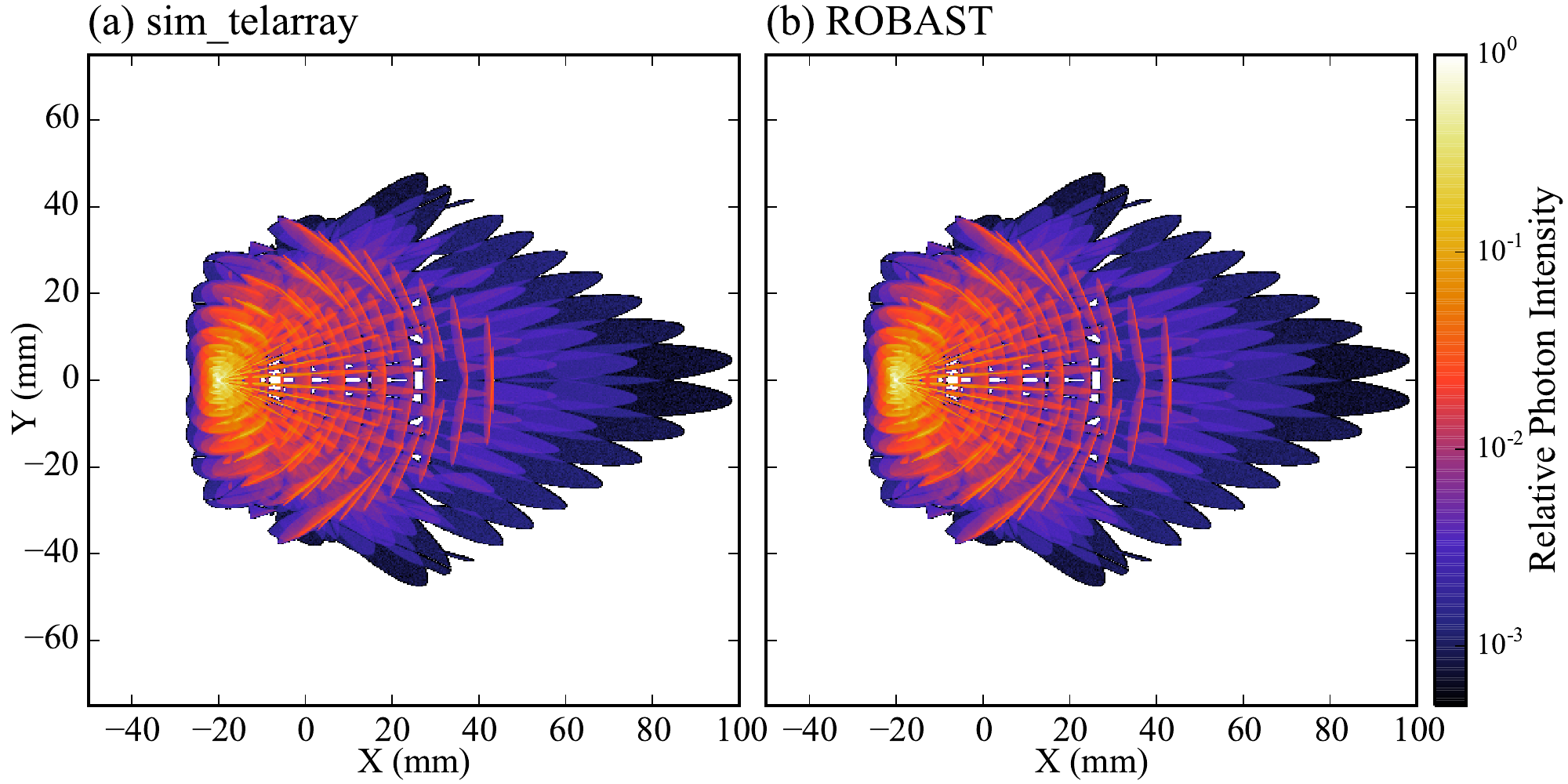}
  \caption{Spot diagrams of the H.E.S.S.~I telescope shown in Figure~\ref{fig:HESS}. (a) Off-axis ($2.5^\circ$) spot simulated with \simtelarray. (b) Same as (a) but simulated with \ROBAST. The spot diagrams shown here are for an ideal H.E.S.S.~I telescope and do not represent the actual performance of the H.E.S.S. telescopes. Misalignment of the segmented mirrors or any deviation in the mirror shapes have not been taken into account.}
  \label{fig:spot_simtelarray}
\end{figure}

Figure~\ref{fig:spot_simtelarray} compares off-axis ($2.5^\circ$) spot diagrams of the H.E.S.S.~I system calculated by \simtelarray\ and \ROBAST, showing an excellent apparent agreement of the two programs. Small spot structures made by the spherical segmented mirrors are visible in the diagrams. One can also see shadowing effect by the telescope structure at around $(X, Y)$ of $(35, 40)$ and $(-15, 35)$, for example. This verifies that rotation and translation matrices used for the alignment of the optics components and the \ROBAST\ photon simulations with the \AMirror\ and \AObscuration\ classes work as expected.

Table~\ref{tab:comparison_simtelarray} gives more quantitative comparison between the two programs. The standard deviations of simulated spot diagrams along $X$ and $Y$ axes were calculated for various field angles, showing consistent values with each other, while very small deviations ($<0.04$\%) up to ${\sim}2$ times larger than the statistical fluctuations are seen.

\begin{table*}
\centering
\caption{Comparison of simulation results by \simtelarray\ and \ROBAST. The standard deviations along $X$ and $Y$ axes are shown for six different field angles $\theta$.}
\label{tab:comparison_simtelarray}
\begin{tabular}{rrrrr}
\toprule
$\theta$ (deg) & \multicolumn{2}{c}{$\sigma_X$ (mm)} & \multicolumn{2}{c}{$\sigma_Y$ (mm)} \\
& \multicolumn{1}{c}{\simtelarray} & \multicolumn{1}{c}{\ROBAST} & \multicolumn{1}{c}{\simtelarray} & \multicolumn{1}{c}{\ROBAST} \\
\midrule
0.0 &  $2.3036\pm0.0003$ & $ 2.3045\pm0.0002$ & $ 2.3072\pm0.0003$ & $ 2.3079\pm0.0002$ \\
0.5 &  $4.4814\pm0.0005$ & $ 4.4816\pm0.0004$ & $ 3.6042\pm0.0004$ & $ 3.6050\pm0.0003$ \\
1.0 &  $8.0860\pm0.0009$ & $ 8.0842\pm0.0008$ & $ 6.0520\pm0.0007$ & $ 6.0502\pm0.0006$ \\
1.5 & $11.9289\pm0.0013$ & $11.9280\pm0.0012$ & $ 8.7362\pm0.0010$ & $ 8.7358\pm0.0009$ \\
2.0 & $16.1769\pm0.0018$ & $16.1792\pm0.0016$ & $11.5113\pm0.0013$ & $11.5090\pm0.0011$ \\
2.5 & $20.6585\pm0.0023$ & $20.6565\pm0.0020$ & $14.2662\pm0.0016$ & $14.2630\pm0.0014$ \\
\bottomrule
\end{tabular}
\end{table*}

\begin{figure}
  \centering
  \includegraphics[width=9cm]{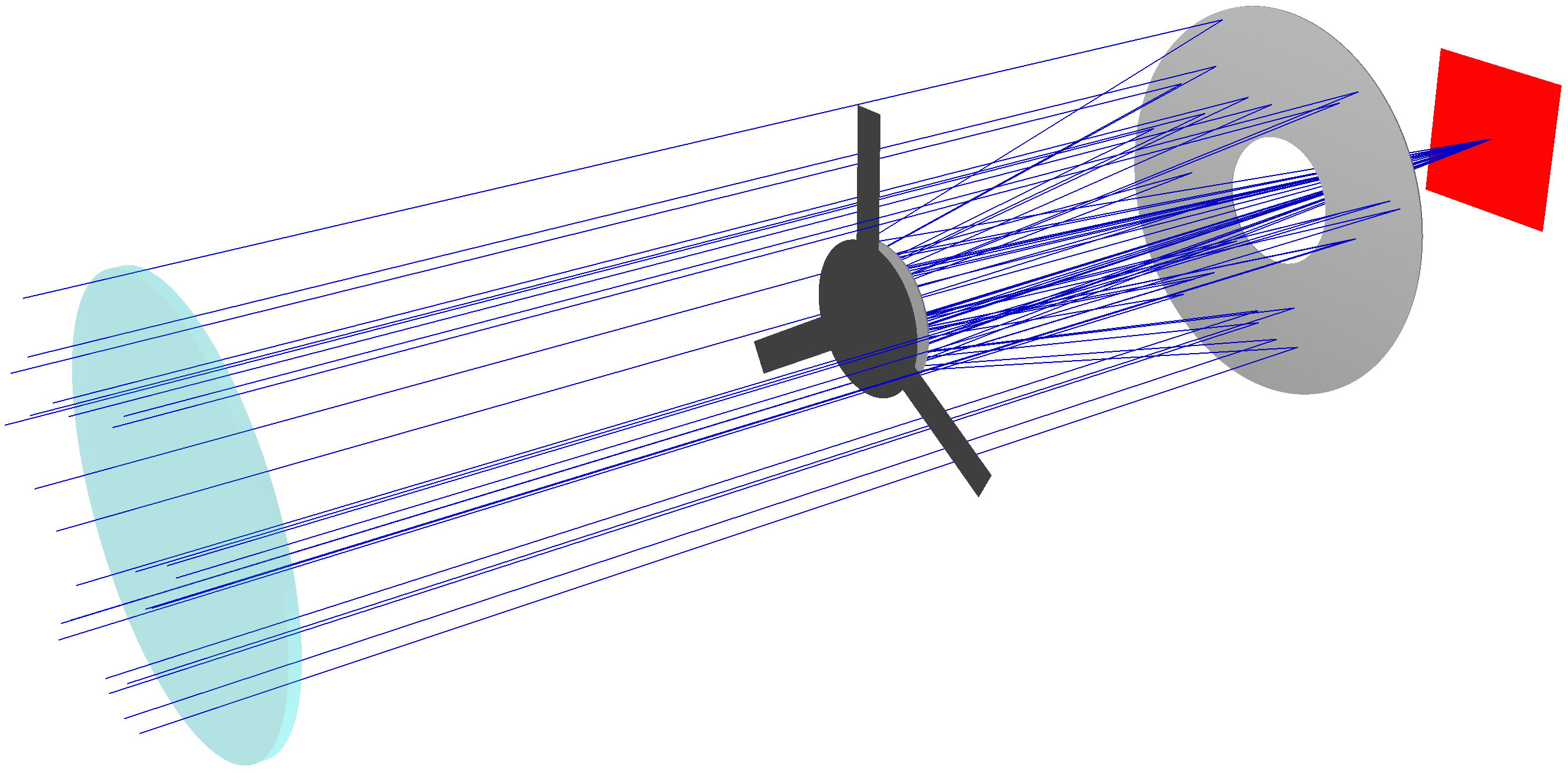}
  \caption{The 3D geometry of a Schmidt--Cassegrain telescope simulated with \ROBAST. The geometry parameters were taken from a \Zemax\ sample file, \texttt{Schmidt-Cassegrain spider obscuration.zmx}, provided in \Zemax. Lines show the tracks of the simulated photons. From left to right, the structures are a plano-aspheric lens, an obscuration comprising three arms and a circular disk, a spherical secondary mirror, an aspherical primary mirror, and a focal plane.}
  \label{fig:spider}
\end{figure}

Figure~\ref{fig:spider} shows the 3D \ROBAST\ geometry of the Schmidt--Cassegrain telescope used for comparison between \Zemax\ and \ROBAST. The plano-aspheric lens, aspherical primary and secondary mirrors, and spider obscuration are modeled with the \ALens, \AMirror, and \AObscuration\ classes, respectively. The geometries of the lens and mirrors are given in the \AGeoAsphericDisk\ class.

\begin{figure}
  \centering
  \includegraphics[width=\columnwidth]{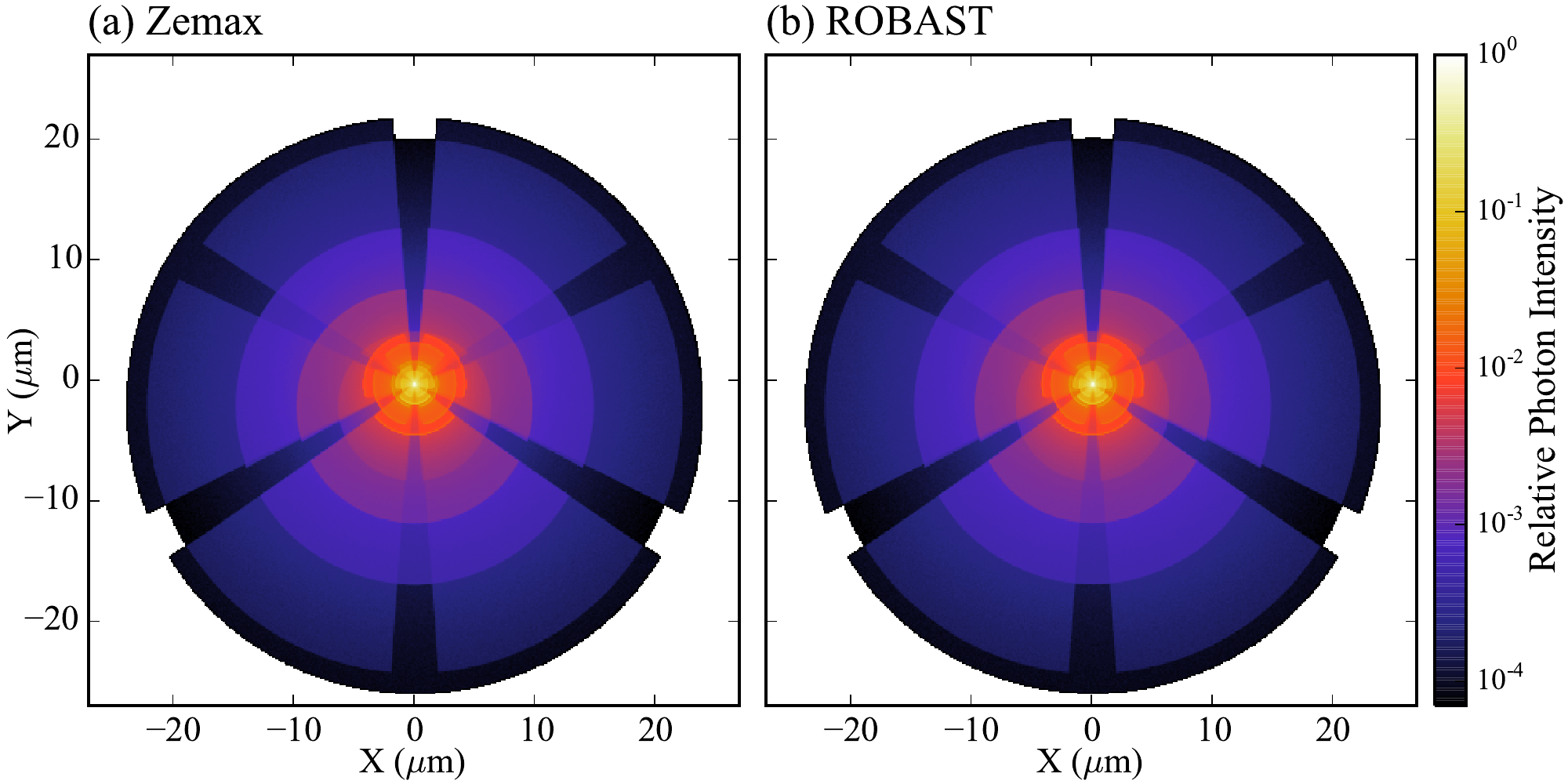}
  \caption{Spot diagrams of the telescope shown in Figure~\ref{fig:spider}. (a) Off-axis ($0.1^\circ$) spot simulated with \Zemax. (b) Same as (a) but simulated with \ROBAST. Note that diffraction is not taken into account in these simulations.}
  \label{fig:spot}
\end{figure}

Figure~\ref{fig:spot} compares the off-axis spot diagrams simulated by \Zemax\ and \ROBAST. The field angle is $0.1^\circ$ and five different wavelengths are present together: $486.1$~nm, 530~nm, 587.6~nm, 610~nm, and 656.3~nm. The two diagrams are consistent, and both programs accurately reproduce the shadows cast by the spider arms, the different spot sizes of the five wavelengths, and the PSFs from the central peak position to the broad tail. This verifies that \ALens, \AMirror, \AObscuration, and \AGeoAsphericDisk\ yield the expected refractions, reflections, obscurations, and geometries, respectively.

In the same way as the cross-check with \simtelarray, Table~\ref{tab:comparison_Zemax} compares the standard deviations of spot diagrams simulated by \Zemax\ and \ROBAST. The spot size difference between the two programs is less than $0.1$\% (${\sim}10$~nm) and it is ${\sim}50$ times smaller than the simulated wavelengths, whereas the difference is significantly larger than the statistical fluctuation. The systematic difference can be ignored in practical applications.

\begin{table*}
\centering
\caption{Comparison of simulation results by \Zemax\ and \ROBAST. The standard deviations along $X$ and $Y$ axes are shown for six different field angles $\theta$.}
\label{tab:comparison_Zemax}
\begin{tabular}{rrrrr}
\toprule
$\theta$ (deg) & \multicolumn{2}{c}{$\sigma_X$ ($\mu$m)} & \multicolumn{2}{c}{$\sigma_Y$ ($\mu$m)} \\
& \multicolumn{1}{c}{\Zemax} & \multicolumn{1}{c}{\ROBAST} & \multicolumn{1}{c}{\Zemax} & \multicolumn{1}{c}{\ROBAST} \\
\midrule
0.0 & $ 5.3321\pm 0.0001$ & $ 5.3297\pm 0.0003$ & $ 5.3323\pm 0.0001$ & $ 5.3300\pm 0.0003$ \\
0.1 & $ 5.4908\pm 0.0001$ & $ 5.4870\pm 0.0003$ & $ 5.5292\pm 0.0001$ & $ 5.5248\pm 0.0003$ \\
0.2 & $ 6.4047\pm 0.0002$ & $ 6.3978\pm 0.0004$ & $ 6.3402\pm 0.0002$ & $ 6.3327\pm 0.0004$ \\
0.3 & $ 8.8677\pm 0.0002$ & $ 8.8595\pm 0.0005$ & $ 8.7167\pm 0.0002$ & $ 8.7074\pm 0.0005$ \\
0.4 & $13.1378\pm 0.0004$ & $13.1267\pm 0.0008$ & $12.8005\pm 0.0003$ & $12.7896\pm 0.0008$ \\
0.5 & $19.0825\pm 0.0005$ & $19.0716\pm 0.0012$ & $18.3659\pm 0.0005$ & $18.3547\pm 0.0011$ \\
\bottomrule
\end{tabular}
\end{table*}

\subsection{Functionality Comparison with Other Programs}
\label{subsec:comparison}

As described in previous sections, \ROBAST\ supports most of the functionality required for CR telescopes. However, other programs such as \Zemax, \simtelarray, and \Geant\ also play important roles in particular situations. Here, for the benefit of users, we compare the functionalities of \ROBAST\ and these existing programs.

\subsubsection{\Zemax}

\Zemax\ is a commercial Windows application which has powerful functionalities for designs and studies of optics and illumination. Among its wide variety of functionalities, the optimization functionality of optics parameters (e.g., lens and mirror shapes) is the most important for CR telescope designs. \ROBAST\ is also able to optimize optics parameters using optimization engines provided with \ROOT, but the user needs to write \CPP\ code for optimization, while \Zemax\ offers an easy graphical user interface.

If one simulates an optical system whose imaging performance is affected by diffraction, it is recommended to use \Zemax\ rather than \ROBAST. However, this is not the case in most CR telescopes. 

\subsubsection{\Geant}
\Geant\ is widely used in simulations of particle physics experiments. It mainly simulates particle interactions and accompanying physics processes as well as tracking of high-energy particles and optical photons. It can also be used for CR telescope simulations as done in the fluorescence detectors of the Pierre Auger Observatory \cite{Abraham:2010:The-fluorescence-detector-of-the-Pierre-Auger-Obse}, but its built-in geometry shapes are limited (see Table~\ref{tab:comparison} for comparisons). Furthermore, \ROOT\ is more widely used in CR telescopes than \Geant, and thus \ROBAST\ is easier to get started with and suitable for those who are already familiar with \ROOT.

\subsubsection{\simtelarray}
\simtelarray\ has been developed to do full simulations of VHE gamma-ray events from telescope to electronics simulations, thus it is not a program dedicated only toward ray-tracing simulations. Users that need a full simulation suite for ground-based gamma-ray telescopes are encouraged to use \simtelarray.

 Its ray-tracing simulation code is faster than \ROBAST, because it uses the sequential method. However, the various optical designs that can be simulated with \simtelarray\ is limited. For example, optical systems with lenses or Winston cones cannot be handled in \simtelarray.

In practice, both \ROBAST\ and \simtelarray\ are used in simulations of CTA. A good example is a shadowing factor parameter that is fed to a configuration file of \simtelarray. In the CTA simulation chain, shadows cast by telescope masts are not fully simulated in \simtelarray. Indeed the shadowing factors of two of the telescope designs proposed for CTA (SCT and GCT, see Sections \ref{subsec:SCT} and \ref{subsec:SST}) are dependent upon field angles. The shadowing factors for these telescopes are calculated in advance using \ROBAST, and the results are written into \simtelarray\ configuration files.

\section{Applications in the Cherenkov Telescope Array}
\label{sec:applications}
\subsection{The Cherenkov Telescope Array}
\label{subsec:CTA}
The Cherenkov Telescope Array (CTA) is a next-generation ground-based gamma-ray observatory comprising two separate arrays of Cherenkov telescopes. These arrays will be constructed in the northern and southern hemispheres, enabling whole sky coverage. The CTA is designed to have a wide energy coverage ($20$~GeV -- $300$~TeV, and its gamma-ray detection sensitivity in the core energy band ($100$~GeV -- $10$~TeV) is expected to be greater by a factor of $10$ than that achievable by the current generation of gamma-ray telescopes \cite{Actis:2011:Design-concepts-for-the-Cherenkov-Telesc,Acharya:2013:Introducing-the-CTA-concept}: H.E.S.S.\footnote{http://www.mpi-hd.mpg.de/hfm/HESS/}, MAGIC\footnote{https://magic.mpp.mpg.de/}, and VERITAS\footnote{http://veritas.sao.arizona.edu/}.

\begin{table*}
\centering
\caption{Telescope Specifications of CTA}
\label{tab:optics}
\begin{tabular}{lccccc}
\toprule
     & Energy Coverage       & Optics Design(s)     & Diameter (m) & Focal Length (m)  & FOV (deg) \\ \hline
LSTs & $20$--$200$~GeV   & Parabola$^\dagger$             & $23$         & $28$              & $4.5$   \\
MSTs & $100$~GeV -- $10$~TeV & Davies--Cotton (DC)$^\ddagger$       & $12$         & $16$              & $7$     \\
SCTs & $200$~GeV -- $10$~TeV & Schwarzschild--Couder (SC) & $9.7$        & $5.6$             & $8$     \\
SSTs & $5$--$300$~TeV  & 2 SC and 1 DC        & $4$      & ${\sim}2.2$ and $5.6$ & ${\sim}9$ \\
\bottomrule
\multicolumn{6}{l}{$^\dagger$ {The parabolic dish shape is approximated by spherical segmented mirrors.}} \\
\multicolumn{6}{l}{$^\ddagger$ {A variant mirror alignment compromises between the time spread and the off-axis PSF.}} \\
\end{tabular}
\end{table*}

\begin{figure*}
  \begin{minipage}[t][0.38\hsize][t]{0.33\hsize}
    (a)
    \begin{center}
      \includegraphics[height=.95\hsize]{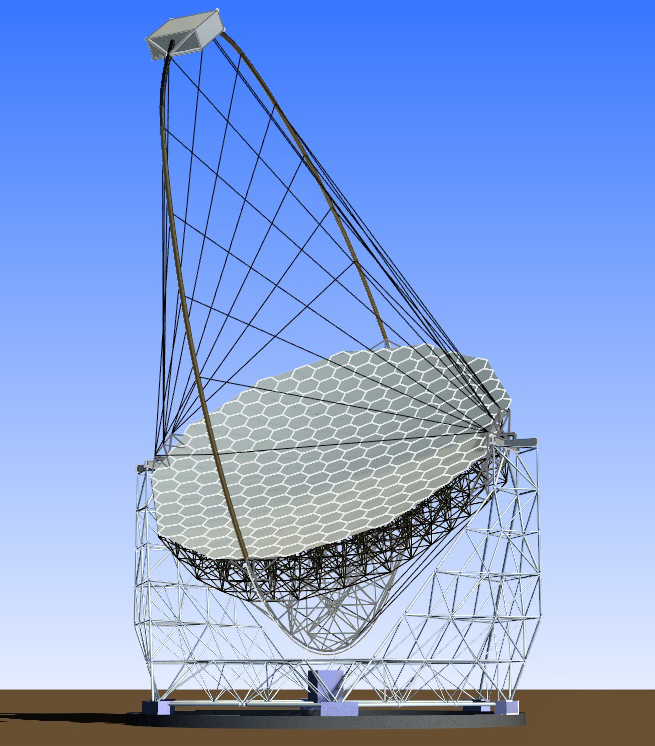}
    \end{center}
  \end{minipage}
  \begin{minipage}[t][0.38\hsize][t]{0.33\hsize}
    (b)
    \begin{center}
      \includegraphics[height=.95\hsize]{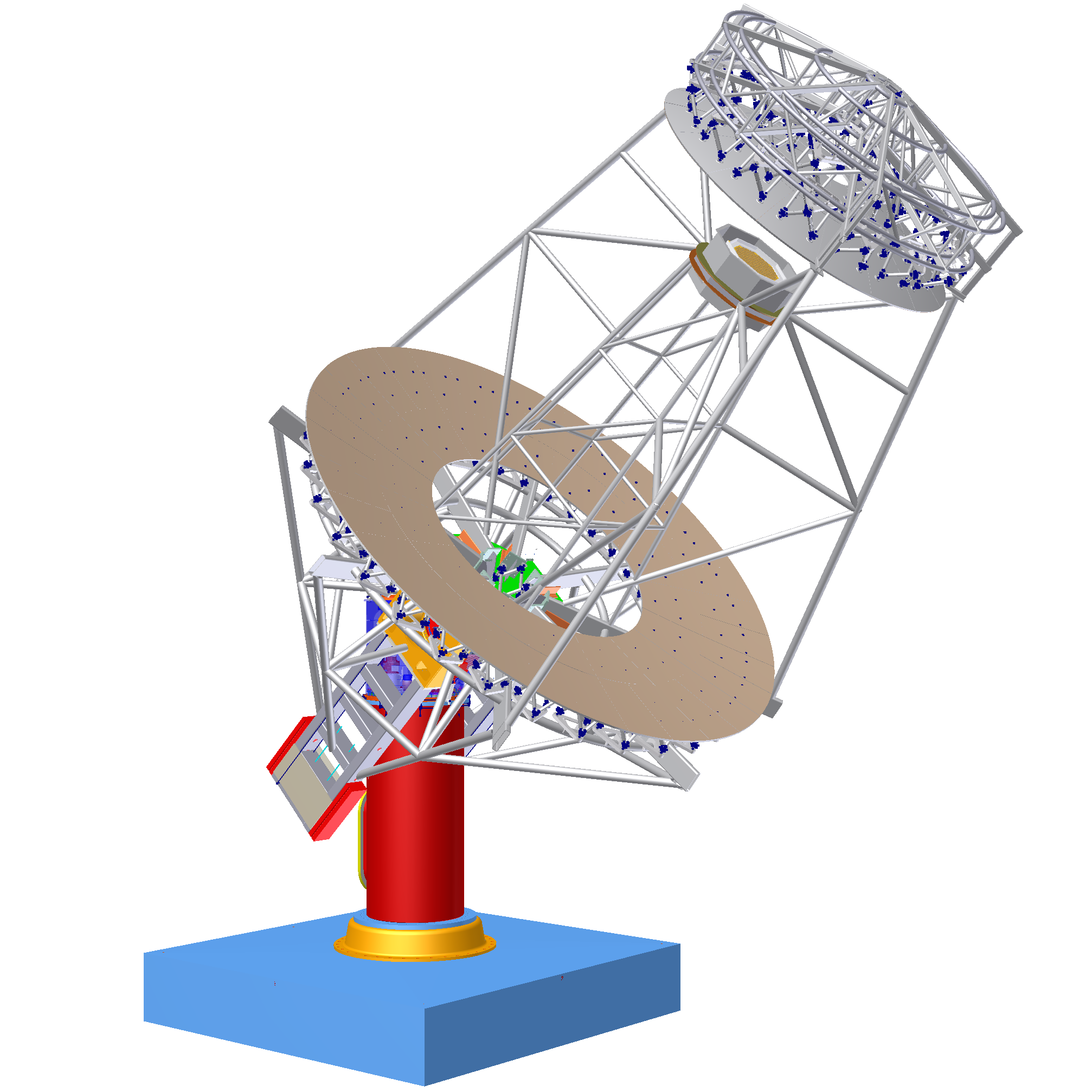}
    \end{center}
  \end{minipage}
  \begin{minipage}[t][0.38\hsize][t]{0.33\hsize}
    (c)\\ \\ \\
    \begin{center}
      \includegraphics[width=\hsize]{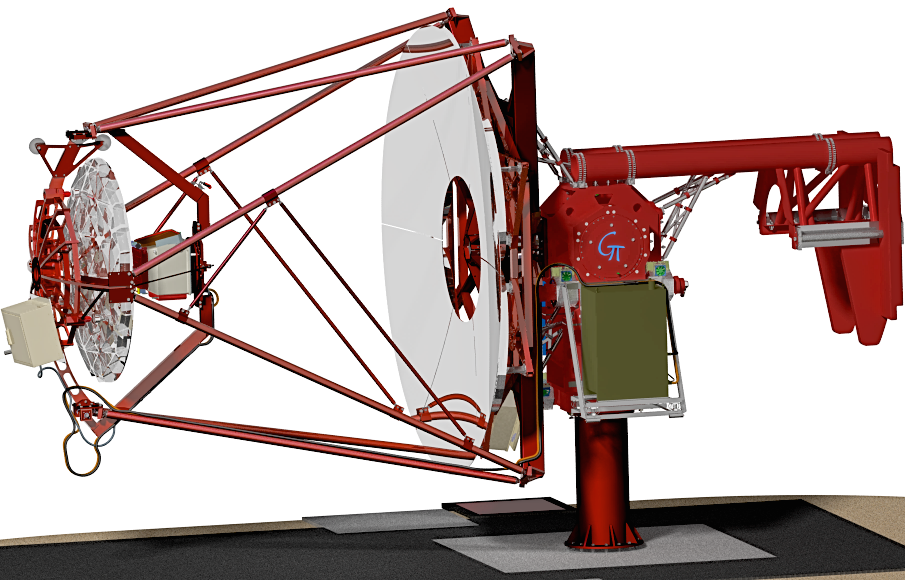}
    \end{center}
  \end{minipage}

  \begin{minipage}[t][0.38\hsize][t]{0.33\hsize}
    (d)
    \begin{center}
      \includegraphics[height=.98\hsize]{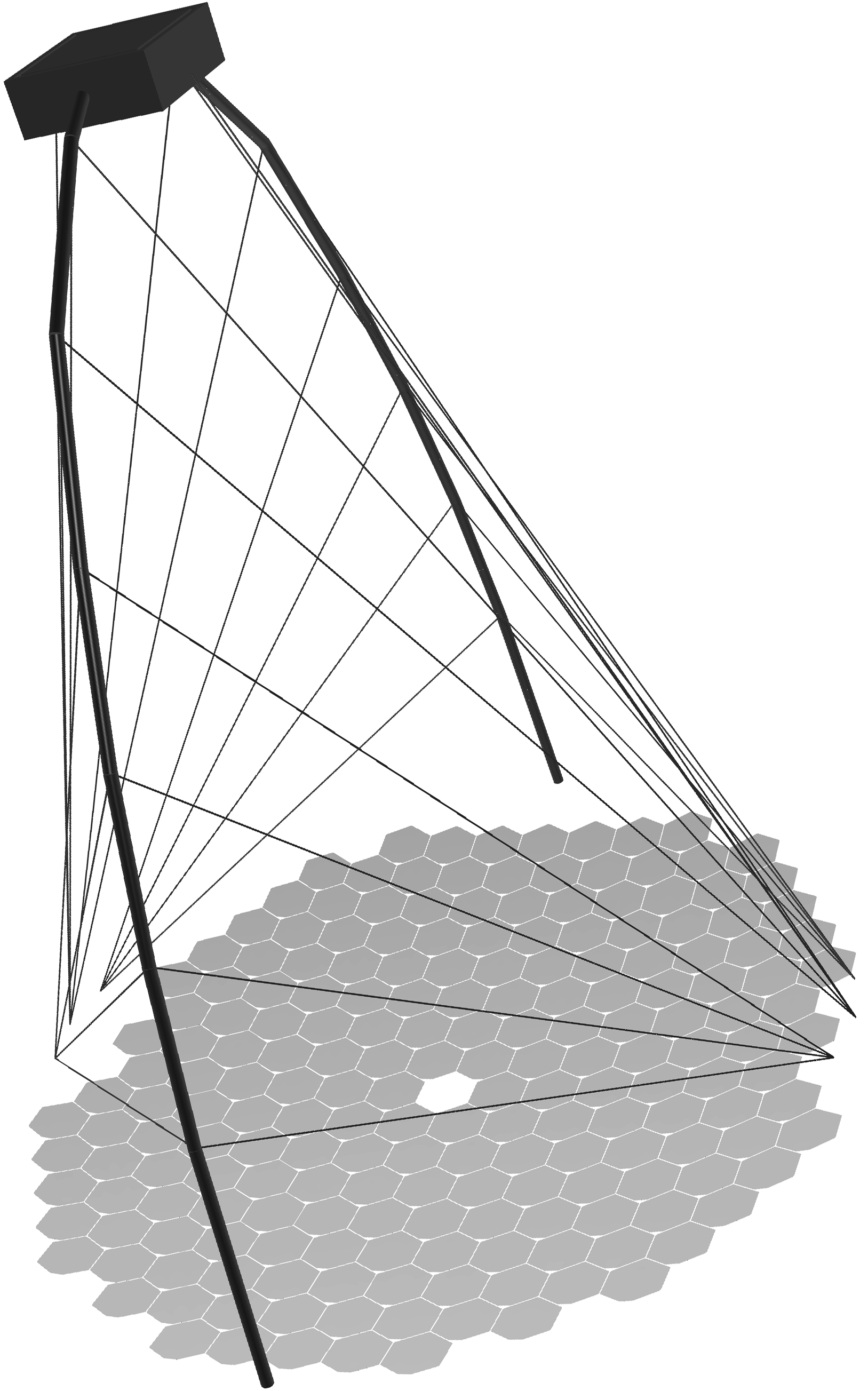}
    \end{center}
  \end{minipage}
  \begin{minipage}[t][0.38\hsize][t]{0.33\hsize}
    (e)
    \begin{center}
      \includegraphics[height=.98\hsize]{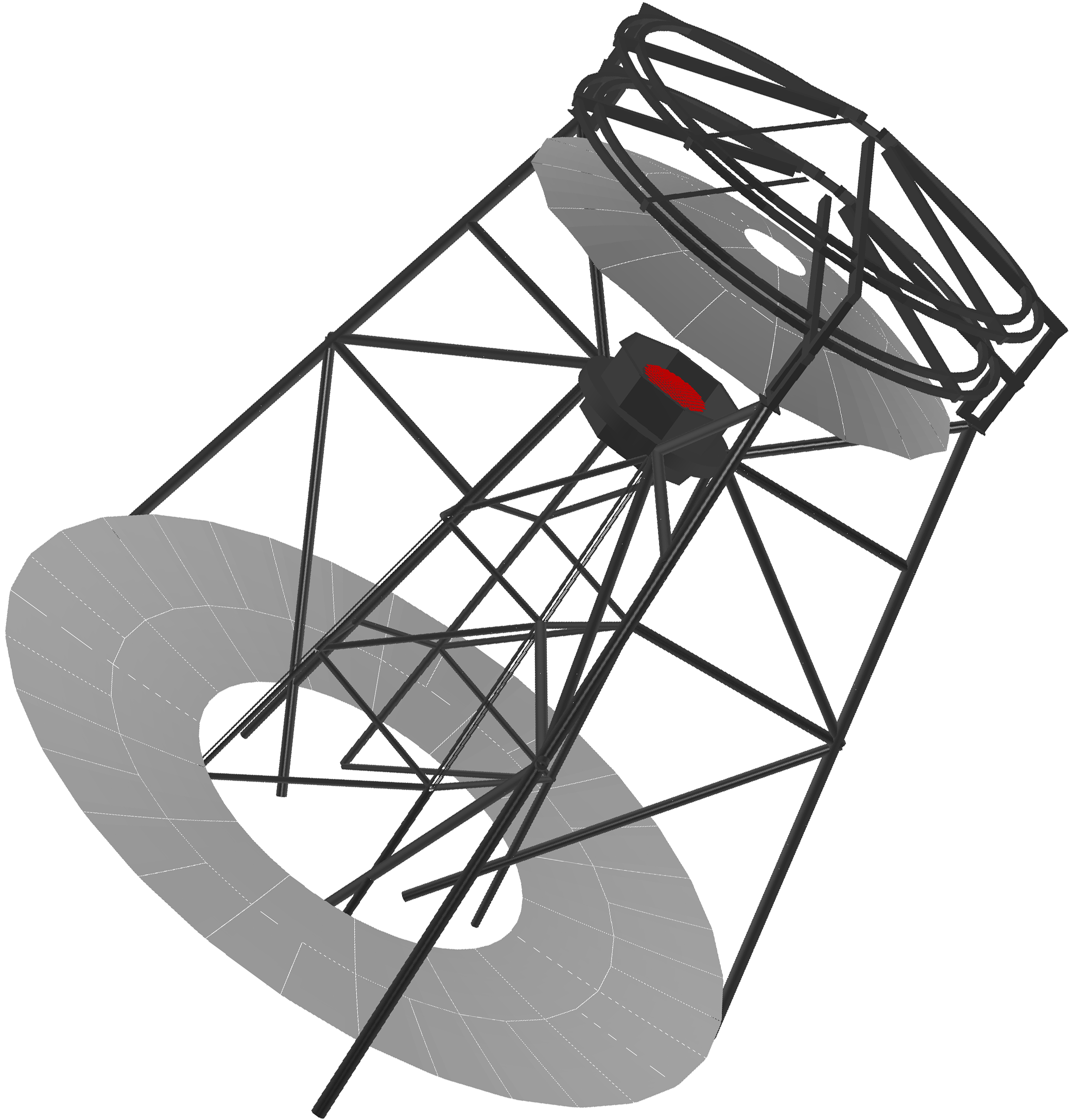}
    \end{center}
  \end{minipage}
  \begin{minipage}[t][0.38\hsize][t]{0.33\hsize}
    (f)
    \begin{center}
      \includegraphics[height=.8\hsize]{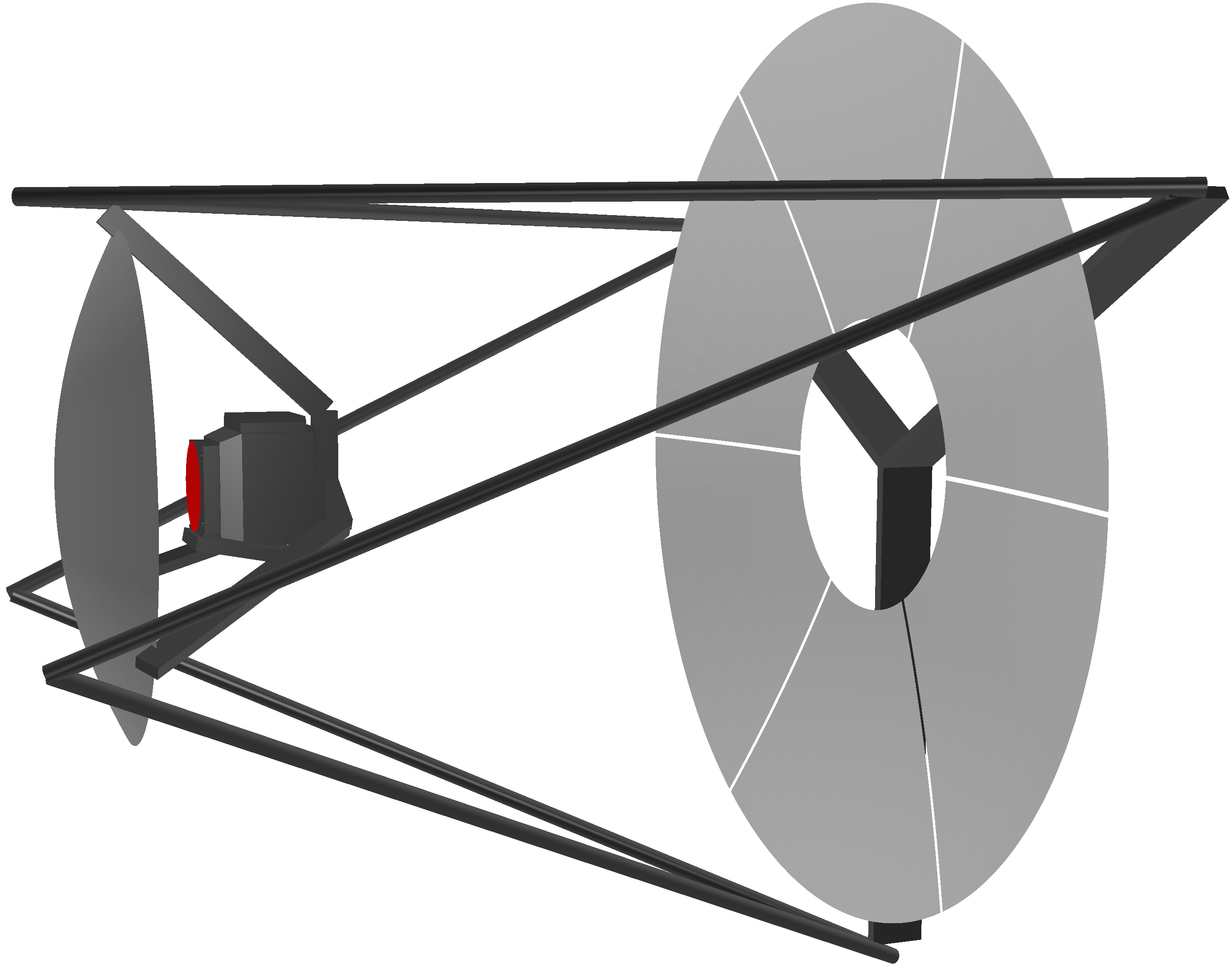}
    \end{center}
  \end{minipage}

  \begin{minipage}[t][0.32\hsize][t]{0.33\hsize}
    (g)
    \begin{center}
      \includegraphics[width=0.98\hsize]{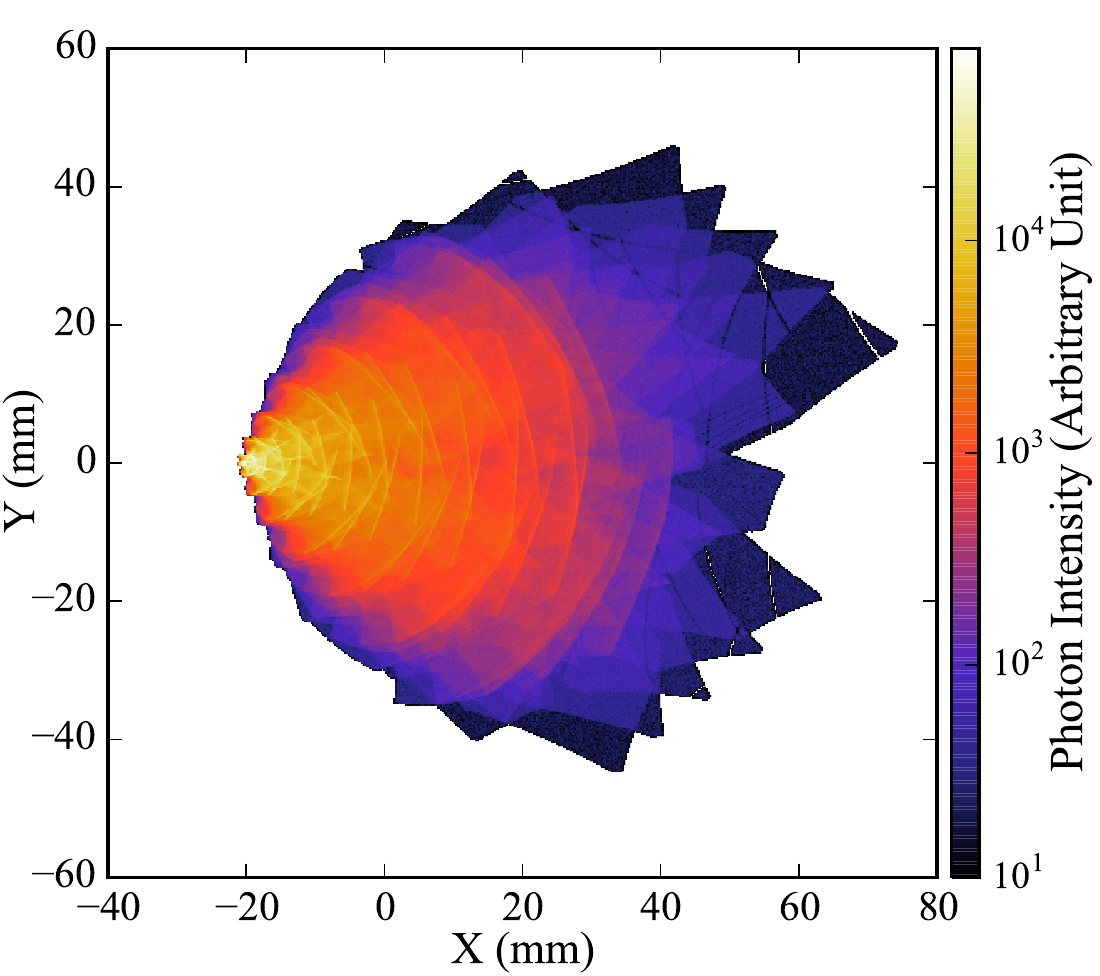}
    \end{center}
  \end{minipage}
  \begin{minipage}[t][0.32\hsize][t]{0.33\hsize}
    (h)
    \begin{center}
      \includegraphics[width=0.98\hsize]{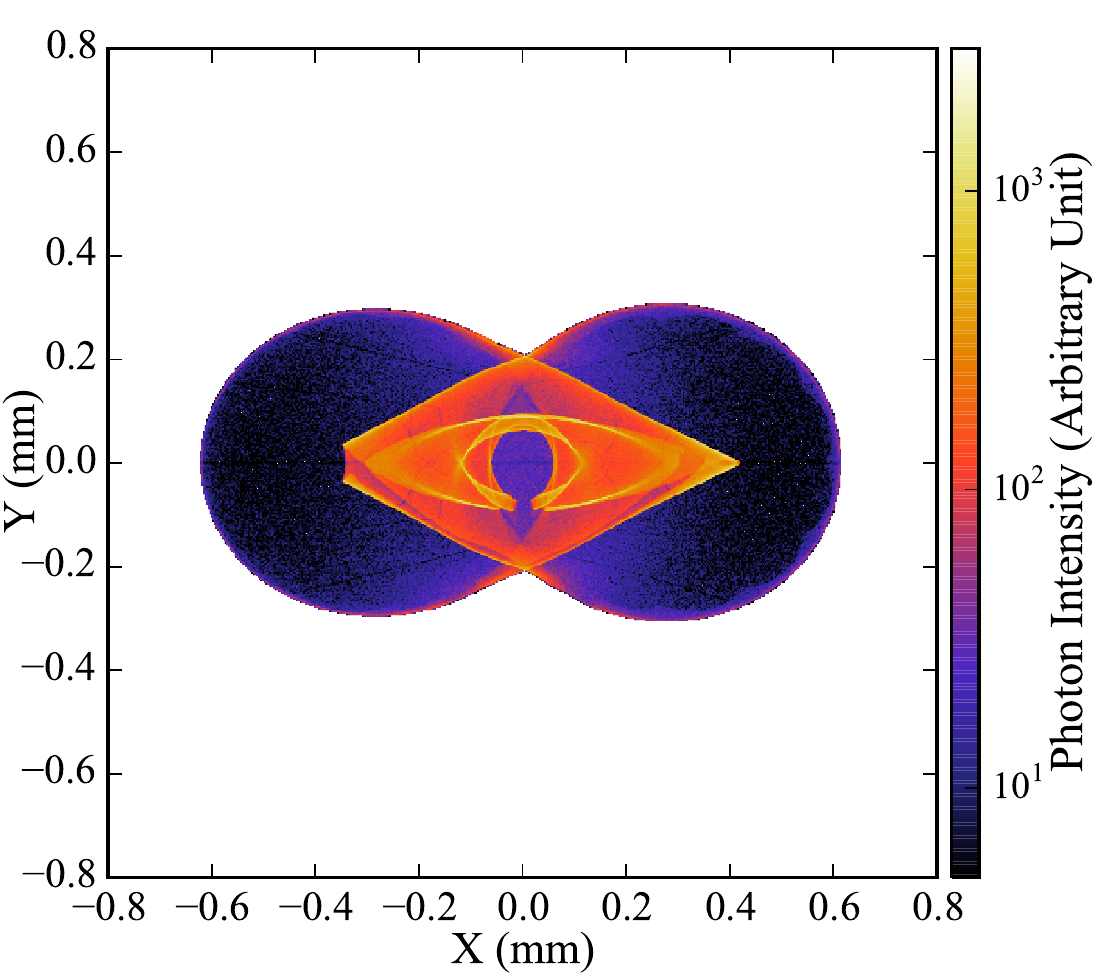}
    \end{center}
  \end{minipage}
  \begin{minipage}[t][0.32\hsize][t]{0.33\hsize}
    (i)
    \begin{center}
      \includegraphics[width=0.98\hsize]{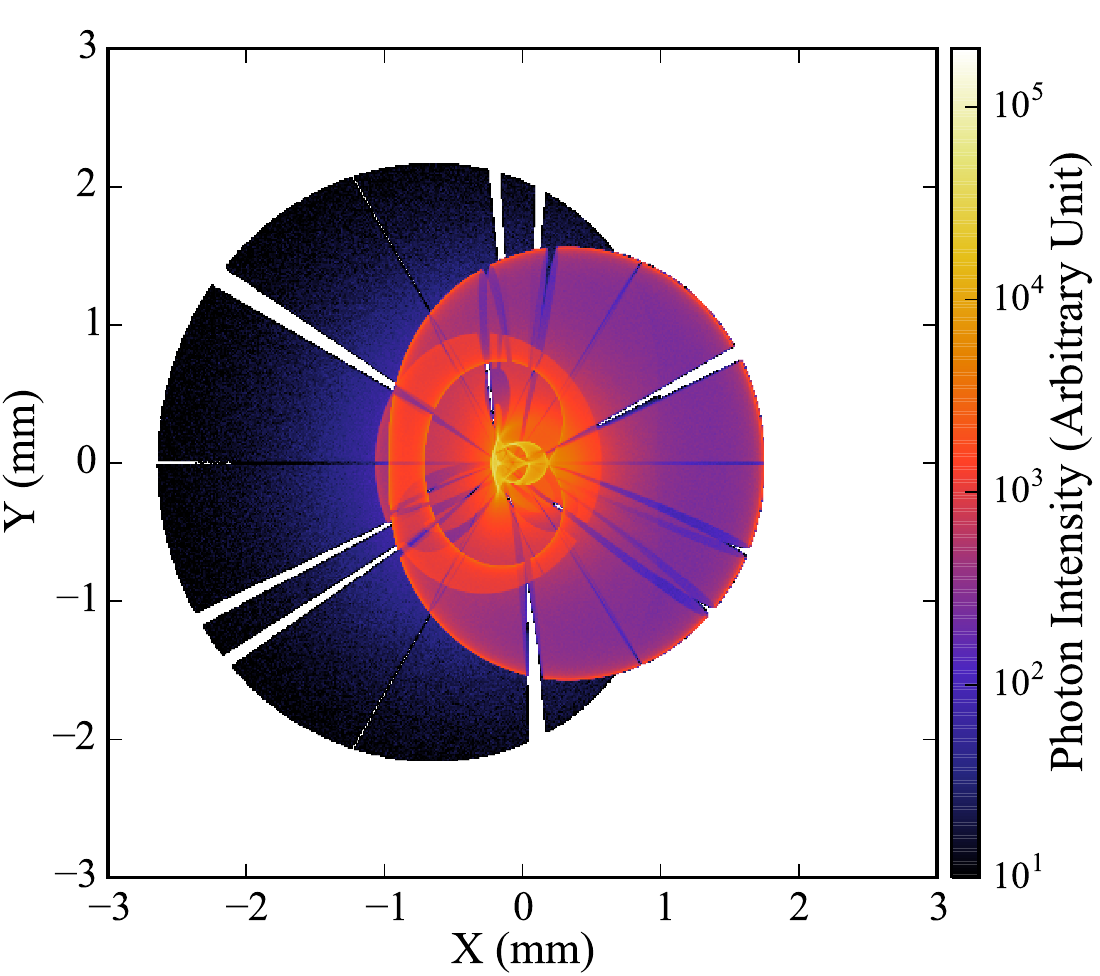}
    \end{center}
  \end{minipage}

  \caption{(a) 3D model of an LST. The camera housing at the upper left is supported by egg-shaped masts which are stiffened with 26 tension ropes. (b) 3D model of an SCT. The large ring-shaped disk is the primary mirror. The octagonal prism box is the camera housing in which 177 camera modules are installed. The upper-right part shows the secondary mirror and its supporting structures. (c) 3D model of a GCT. Located in sequence from left to right are the secondary mirror, the camera housing, and the primary mirror similarly to the SCT optical system. Note that the final designs and mirror positions of the three telescopes may differ from those of the 3D models presented here. (d)--(f) Equivalent \ROBAST\ geometries of (a)--(c), showing only the components casting shadows on the segmented mirrors. For visualization purposes, the thin ropes in (d) appear wider than in reality. (g)--(i) Spot diagrams of the optical systems at a field angle of $1^\circ$. The spot diagrams shown here are for ideal optical systems and do not represent the actual performance. Misalignment of the segmented mirrors or any deviation in the mirror shapes have not been taken into account. (Image credit for figures (a)--(c): the CTA Consortium. (d)--(i): Taken from \cite{Okumura:2015:ROBAST:-Development-of-a-Non-Sequential-Ray-Tracin})}
    \label{fig:telescopes}
\end{figure*}

Several telescope designs that simultaneously achieve a wide effective area, high angular resolution, and wide energy coverage have been proposed and are listed in Table~\ref{tab:optics}. Large-sized telescopes (LSTs) cover the lowest energy band ($20$--$200$~GeV) with the largest-diameter mirror comprising segmented spherical mirrors aligned in a parabolic shape, as illustrated in Figure~\ref{fig:telescopes}(a).

The middle energy band ($100$~GeV -- $10$~TeV) is mainly observed by medium-sized telescopes (MSTs). By virtue of a variant design of Davies--Cotton (DC) optics, MSTs realize a wider field-of-view (FOV) with more uniform angular resolution over the FOV than parabolic telescopes. To increase the effective area and improve the angular resolution of gamma-ray arrival directions in this energy band, Schwarzschild--Couder Telescopes (SCTs, also referred to as SC-MSTs) have been proposed as an extension of DC MSTs in the southern array. As their name indicates, SCTs are installed with an SC optics design comprising aspherical primary and secondary mirrors. Whereas the angular resolution of DC MSTs is ${\sim}10$-arcmin, that of an SCT is improved to ${\sim}4$~arcmin. The mirrors in the SC design will be composed of 72 segmented mirror facets, as illustrated in Figure~\ref{fig:telescopes}(b).

The highest energy band ($5$--$300$~TeV) is covered by small-sized telescopes (SSTs). Three SST telescope designs (two SC and one DC) have been proposed. One of these is referred to as the Gamma Cherenkov Telescope (GCT; formerly SST-GATE) which also simultaneously achieves a wide FOV and high angular resolution using an SC optics system. Figure~\ref{fig:telescopes}(c) shows a 3D CAD image of a GCT.

The northern array will constitute $4$ LSTs and ${\sim}15$ MSTs. This array will mainly observe low-energy gamma rays from extragalactic objects such as active galactic nuclei and gamma-ray bursts. In contrast, the southern sky will be observed by $4$ LSTs, ${\sim}25$ MSTs, and $70$--$90$ SSTs (additional ${\sim}25$ SCTs may be deployed), enabling a survey of the inner Galactic plane with higher sensitivity and a wider energy coverage (up to $300$~TeV) than is currently possible.

Among the six telescope designs, we used the \ROBAST\ library for the LSTs, SCTs, and SSTs (GCTs). Using common software for the different CTA sub-projects, we substantially reduced the time and effort of software development in the large global collaboration. Moreover, the ROBAST functionality described in Section~\ref{subsec:requirements} is quite suitable for CTA telescopes comprising a large number of segmented mirrors and telescope frames. Such telescopes are important in image simulations, evaluation of shadowing, and tolerance analysis. Telescope array simulations with \ROBAST\ are also integrated with other programs; e.g., \GrOptics/\CARE\footnote{http://otte.gatech.edu/care/} adopts \ROBAST\ as their ray-tracing engine. In the following subsections, we demonstrate the actual use of ROBAST in four CTA applications.

\subsection{Large-Sized Telescopes}
\label{subsec:LST}
An LST has 198 segmented mirrors aligned on a parabolic surface. The whole surface diameter is $23$~m and the focal length is $28$~m \cite{Ambrosi:2013:The-Cherenkov-Telescope-Array-Large-Size-Telescope}. Each mirror facet has a spherical surface with a radius of curvature of ${\sim}56$--$58.4$~m to approximate the parabolic dish. Its outer shape is a regular hexagon with a diameter of $1.51$~m (side to side), but one of its vertices is sliced off to make space for an alignment-monitoring CMOS camera. The shape of the mirror facet is non-trivial, but can be constructed from Boolean operations in \libGeom\ (as explained in Figure~\ref{fig:composite}). Figure~\ref{fig:telescopes}(d) shows the LST geometry built with \ROBAST. The segmented mirrors, camera housing, telescope masts, and 26 tension ropes are accurately reproduced.

Figure~\ref{fig:telescopes}(g) presents a \ROBAST\ simulation of the LST optical system at a field angle of $1^\circ$. Since the LST dish is nearly parabolic, comatic aberration caused by the outer segmented mirrors appears in the spot diagram. The diagram is vertically asymmetric, reflecting the asymmetry in the positions of the $198$ segmented mirrors. Many small structures introduced by the facet shape are also visible. Thin shadows cast by the tension ropes are visible on the outlying photons, verifying that the telescope structure built with the \AObscuration\ class properly casts shadows on the mirrors.

As one might expect, Davies--Cotton optical systems are easily simulated by the same approach. Davies--Cotton system can be constructed from hexagonal segmented mirrors, and their arbitrary positions and directions can be given using \libGeom.

\subsection{Schwarzschild--Couder Telescopes}
\label{subsec:SCT}
An SCT comprises aspherical primary and secondary mirrors, which suppress the aberrations appearing in wide FOV optical systems \cite{Vassiliev:2007:Wide-field-aplanatic-two-mirro,Rousselle:2013:Schwarzschild-Couder-telescope-for-the-Cherenkov-T2,Vassiliev:2013:Schwarzschild-Couder-telescope-for-the-Cherenkov-T}. The angular resolution of the SCT is ${\lesssim}4$~arcmin at field angles up to $4^\circ$, which cannot be achieved by conventional parabolic or Davies--Cotton telescopes with a similar $f/D$ ratio. However, this optical system is built from a large number of aspherical segmented mirrors and includes a large-diameter ($5.4$~m) secondary mirror with supporting structures (Figure~\ref{fig:telescopes}(b)). Such a complex optical system with an accurate 3D geometry is difficult to simulate with \Geant\ or other software using the sequential method.

Here, we successfully constructed the complex SCT geometry using \ROBAST\ as shown in Figure~\ref{fig:telescopes}(e). Tetragonal and pentagonal segmented mirrors with aspherical surfaces, the focal plane comprising $177$ camera modules ($11,328$ image pixels), and the telescope frames were accurately reproduced.

Figure~\ref{fig:telescopes}(h) shows a spot diagram of the SCT optical system at a field angle of $1^\circ$. The diagram is almost vertically symmetric, but asymmetric shadows cast by the telescope masts can been seen. Very thin shadows cast by small gaps between segmented mirrors are also visible.

\subsection{Small-Sized Telescopes}
\label{subsec:SST}

The GCT optical system, one of the three SST design proposals \cite{Laporte:2012:SST-GATE:-an-innovative-telescope-for-very-high-en, Zech:2013:SST-GATE:-A-dual-mirror-telescope-for-the-Cherenko}, is very similar to the SCT design, because it also employs an SC optical system. However, the primary mirror of a GCT comprises only 6 segmented mirrors, and the secondary is monolithic rather than segmented (see Figure~\ref{fig:telescopes}(c)). The secondary mirror and camera housing in the GCT design are supported by tubular masts and rectangular trusses, respectively. These structures cast shadows on the mirror surfaces.

In GCT development, \ROBAST\ was used to accurately calculate the GCT effective area and to evaluate the mirror misalignment tolerance \cite{Rulten:2015}. Panels (f) and (i) of Figure~\ref{fig:telescopes} show \ROBAST\ simulations of the GCT optical system and its off-axis PSF, respectively\footnote{The \ROBAST\ simulations were performed on an old telescope design that differs from Figure~\ref{fig:telescopes}(c).}. Shadows cast by the telescope masts are clearly visible in Figure~\ref{fig:telescopes}(i).

\subsection{Hexagonal Light Concentrators}
\label{subsec:LG}
\begin{figure}
  \centering
  \includegraphics[width=6cm]{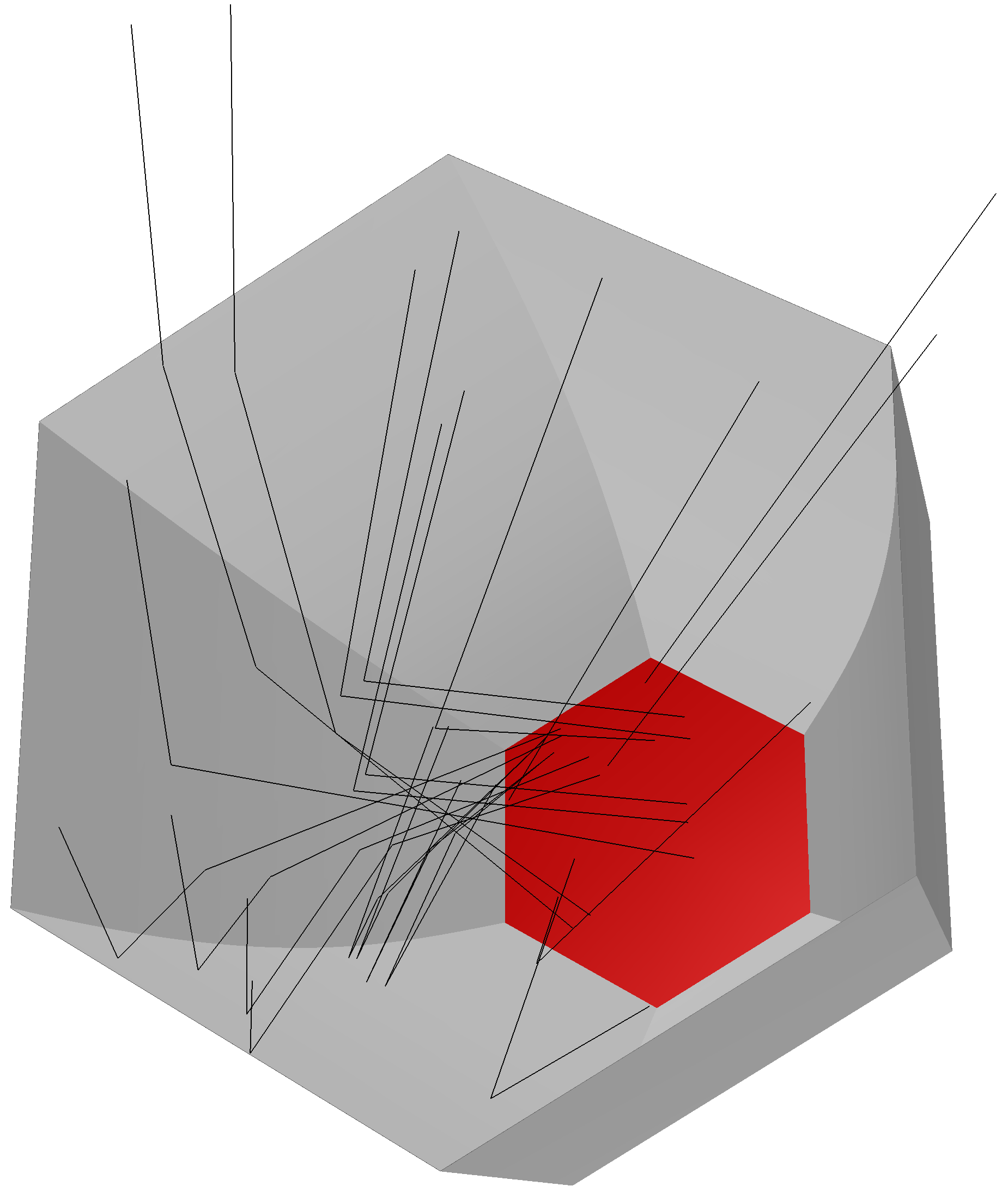}
  \caption{A simulation of a hexagonal Okumura cone with a cutoff angle of $30^\circ$. The hexagonal tapered geometry (light gray in the online version) represents the reflective surfaces of the cone modeled with \AMirror, and the hexagon at the bottom of the cone is a photodetector (red, \AFocalSurface). Polylines represent photon tracks. This figure was taken from \cite{Okumura:2015:ROBAST:-Development-of-a-Non-Sequential-Ray-Tracin}.}
  \label{fig:Okumura}
\end{figure}

\begin{figure}
  \centering
  \includegraphics[width=8cm]{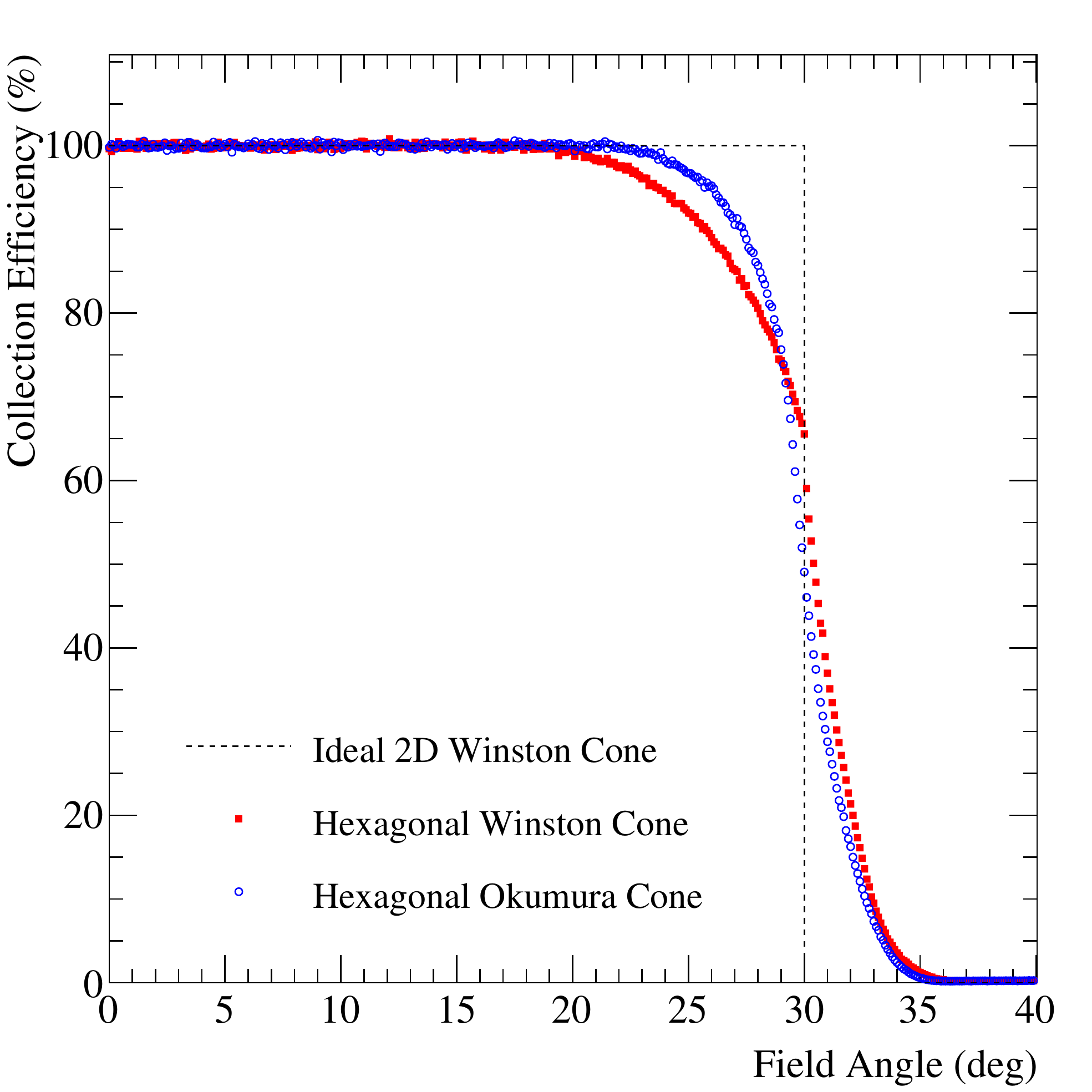}
  \caption{Comparison of collection efficiencies of an ideal 2D Winston cone (dashed line), hexagonal Winston cone (filled red squares), and hexagonal Okumura cone (open blue circles). The cutoff angle for all cones is $30^\circ$. This figure was taken from \cite{Okumura:2015:ROBAST:-Development-of-a-Non-Sequential-Ray-Tracin}.}
  \label{fig:collection}
\end{figure}

\ROBAST\ has built-in classes that simulate the hexagonal light concentrators widely used in CR telescopes. This feature is a distinct advantage of \ROBAST. A conventional hexagonal Winston cone, which is composed of six parabolic walls, can be easily simulated using the \AGeoWinstonConePoly\ class\footnote{Axisymmetric and other polygonal cones can be simulated as well.}. A variant light concentrator composed of six \Bezier-curve walls (an Okumura cone \cite{Okumura:2012:Optimization-of-the-collection-efficiency-of-a-hex}) can also be simulated with the \AGeoBezierPgon\ class as shown in Figure~\ref{fig:Okumura}. This simulation reveals extensive non-sequential photon tracking by \ROBAST; moreover, the multiple reflections on the cone surfaces are well-reproduced. Figure~\ref{fig:collection} compares the collection efficiencies of an ideal 2D Winston cone, a hexagonal Winston cone, and a hexagonal Okumura cone. These efficiencies were directly calculated and plotted with the \TGraph\ class of \ROOT.

Currently, a hexagonal light concentrator design proposed for the LST cameras is being developed using \ROBAST. The cone shape was optimized in \ROBAST\ simulations using the optimization scheme developed by \citet{Okumura:2012:Optimization-of-the-collection-efficiency-of-a-hex}. To reproduce the light concentrator performance more accurately, the simulation accounted for the incidence angle and position dependence of the PMT photodetection efficiency, exploiting the functionality of the \AFocalSurface\ class. Consequently, the final simulation result and measured collection efficiency of an LST light guide prototype significantly differs from that shown in Figure~\ref{fig:collection}. In addition, using \ROBAST\ simulations, the UV-enhanced coating of the cone was optimized in terms of the photon incidence angle distribution on its surface.

\section{Conclusion}

We have developed a new \CPP\ library, \ROBAST, which is intended for ray-tracing simulations of CR telescopes. To demonstrate the usefulness of \ROBAST, we successfully simulated three telescope designs as well as a light concentrator proposed for CTA. Through these simulations we were able to confirm that the ROBAST functionalities meet the software requirements for accurate simulation of CR telescopes. In addition, we illustrated that \ROBAST\ was able to simulate complex optical systems as accurately as independently produced programs such as \Zemax, which are available commercially. The \ROBAST\ library is freely available online and is expected to be used for the development of other CR telescopes as well as further simulations of CTA.

\section*{Acknowledgments}
We are grateful to Dr. Andrei Gheata, the principal developer of the \ROOT\ geometry library. We could not have developed \ROBAST\ without his help and support. Dr. Akito Kusaka helped A.~O. develop \CPP\ code and initiated the idea of ray-tracing simulations. We also thank a number of colleagues in the Ashra collaboration and the CTA Consortium, who worked on the telescope optics designs and helped us simulate the \ROBAST\ applications. We gratefully acknowledge Dr. Konrad Bernl\"{o}hr, who wrote \simtelarray\ and kindly allowed us to use part of his code in \ROBAST. This study was supported by JSPS KAKENHI Grant Numbers 25610040 and 25707017. A.~O. was supported by a Grant-in-Aid for JSPS Fellows.





\bibliographystyle{model1-num-names}
\bibliography{oxon}







\end{document}